\documentclass[manuscript]{aastex}

\newcommand{\kms}{\,$\>\mathrm{km\>s}^{-1}\>$}
\def\up#1{\leavevmode \raise.16ex\hbox{#1}}

\slugcomment{Revised Submission: \today}

\shorttitle{TEXES [Fe~II] Observations of M Supergiants}
\shortauthors{Harper et al.}

\begin{document}


\title{TEXES Observations of M Supergiants: \\
Dynamics and Thermodynamics of Wind Acceleration}


\author{Graham M. Harper\altaffilmark{1,2}}
\affil{School of Physics, Trinity College, Dublin 2, Ireland}
\altaffiltext{1}{Center for Astrophysics and Space Astronomy, University of Colorado,
    Boulder, Colorado 80309 USA}
\altaffiltext{2}{Visiting Astronomer at the Infrared Telescope Facility, which is operated by the University of Hawaii under Cooperative Agreement no. NCC 5-538 with the National Aeronautics and Space Administration, Science Mission Directorate, Planetary Astronomy Program.}

\author{Matthew J. Richter\altaffilmark{2}}
\affil{Department of Physics, University of California at Davis, CA 95616}

\author{Nils Ryde}
\affil{Lund Observatory, SE-221 00 Lund, Sweden}

\author{Alexander Brown\altaffilmark{2}}
\affil{Center for Astrophysics and Space Astronomy, University of Colorado,
    Boulder, CO 80309}

\author{Joanna Brown}
\affil{Max-Planck-Institut f\"ur Extraterrestrisches Physik, Germany}

\author{Thomas K. Greathouse\altaffilmark{2}}
\affil{Southwest Research Institute, San Antonio, TX 78228}

\author{Shadrian Strong\altaffilmark{2}}
\affil{Johns Hopkins Applied Physics Lab, Laurel, MD 20723}



\begin{abstract}
We have detected \up[\ion{Fe}{2}\up] 17.94\,$\mu$m and 24.52\,$\mu$m
emission from a sample of M supergiants ($\mu$~Cep, $\alpha$~Sco, $\alpha$~Ori,
CE~Tau, AD~Per, and $\alpha$~Her)
using the Texas Echelon Cross 
Echelle Spectrograph on NASA's Infrared Telescope Facility.
These low opacity emission lines are resolved at $R\simeq
50,000$ and provide new diagnostics of the dynamics and
thermodynamics of the stellar wind acceleration zone. 
The \up[\ion{Fe}{2}\up] lines, from the first excited term ($a\>^4F$), 
are sensitive to the warm plasma where energy is deposited into
the extended atmosphere to form the chromosphere and wind outflow. These diagnostics
complement previous {\it Kuiper Airborne Observatory} and {\it Infrared Satellite Observatory} observations which were sensitive to
the cooler and more extended circumstellar envelopes.
The turbulent velocities of $V_{turb} \simeq 12-13\>{\rm km\>s}^{-1}$ observed in the 
\up[\ion{Fe}{2}\up] $a\>^4F$ forbidden lines are found to be a common property of 
our sample, and are less than that 
derived from the hotter
chromospheric \ion{C}{2}\up] 2325\,\AA {}
lines observed in $\alpha$~Ori, where $V_{turb} \simeq 17-19\>{\rm km\>s}^{-1}$. 
For the first time, we have dynamically resolved the motions of the dominant
cool atmospheric component discovered in $\alpha$~Ori from
multi-wavelength radio interferometry by \cite{lim_etal98}. 
Surprisingly, the emission centroids are quite Gaussian and at rest 
with respect to the M supergiants. These constraints combined with model
calculations of the infrared emission line fluxes for $\alpha$~Ori 
imply that the warm material has a low outflow velocity
and is located close to the star. We have also detected 
narrow \up[\ion{Fe}{1}\up] 24.04\,$\mu$m emission that confirms that \ion{Fe}{2} 
is the dominant ionization state in $\alpha$~Ori's extended atmosphere.
\end{abstract}

%% subject header.  Objects should be in the appropriate "individual"
%% headers (e.g. quasars: individual, stars: individual, etc.) with the
%% additional provision that the total number of headers, including each
%% individual object, not exceed six.  The \objectname{} macro, and its
%% alias \object{}, is used to mark each object.  The macro takes the object
%% name as its primary argument.  This name will appear in the paper
%% and serve as the link's anchor in the electronic edition if the name
%% is recognized by the data centers.  The macro also takes an optional
%% argument in parentheses in cases where the data center identification
%% differs from what is to be printed in the paper.



\section{INTRODUCTION}

M supergiants present a particular challenge in the study of mass-loss from 
cool evolved stars. For the K through mid-M spectral-types
there are no working theories that can satisfactorily explain their observed wind properties. 
It has long been recognized that mass-loss driven by radiation pressure on dust 
does not satisfy the energy-budget requirement for overcoming the gravitational potential
\citep{holzer_macgregor85}. Both indirect evidence from silicate dust temperatures inferred through
semi-empirical modeling, e.g., \cite{david_papoular90}, 
and direct evidence from infrared (IR) interferometry \citep{danchi_etal94}  
show that the inner radius of the dominant dust features are 
located far from the stellar surface ($\sim 5-30R_\ast$), and therefore some other mechanism is
responsible for lifting the material out of the stellar gravitational potential. 
Observations reveal that there is insufficient hot plasma to drive thermal
Parker-type winds. While mass-loss from some form of pulsation or convective 
ejection events has yet to be demonstrated, the winds of M supergiants
often show complex structures. For example, M supergiants
show multiple absorption  in the CO 4.6~$\mu$m fundamental band \citep{bernat81},
and the 12.5\,$\mu$m and 20.8\,$\mu$m images of $\alpha$~Scorpii (M1~Iab + B3~V) show 
that the dust is
clumped \citep{marsh_etal01}.

To drive the observed mass-loss rates ($10^{-7} - 10^{-5}\ {\rm M}_\odot\>{\rm yr}^{-1}$)
some process, or combination of processes, must substantially increase the density scale-height
close to the star above the thermal hydrostatic value.
A promising mass-loss mechanism for K and M stars of luminosity classes III (giants) through I 
(supergiants) emerged in the 1980's in the form of Alfv\'en 
wave-driven winds (\citealt{hartmann_macgregor80, hartmann_avrett84}).  Unlike acoustic waves 
and shocks which dissipate too close to the star, the long dissipation lengths
of the non-compressive MHD waves provide a possible explanation for driving the observed
mass-loss rates. These idealized  Alfv\'en wave-driven wind models (e.g., Wentzel-Kramers-Brillouin approximation) 
also suffer from 
theoretical problems that require fine-tuning of the wave damping length to avoid terminal wind speeds 
in excess of those observed \citep{holzer_etal83}. 
A characteristic of the 1-D Alfv\'en wave-driven models was that they predicted a bloated and turbulent
wind acceleration zone that was also a potential source of copious chromospheric emission that had 
been observed in many evolved K-M stars with the {\it International Ultraviolet Telescope (IUE)}.
The total Alfv\'en energy fluxes and line-widths of the observed ultraviolet (UV) chromospheric emission 
appeared to be in reasonable agreement with the models if the magnetic fields were  0.1-1.0~mT (1-10 Gauss), 
especially if area filling factors were included 
(e.g., \citealt{hartmann_etal81,harper88}).
 
However, observations with spectrographs on board the {\it Hubble Space Telescope (HST)} revealed that this was 
not the case. The higher spectral resolution and higher signal-to-noise ratio spectra revealed that the 
optically thin UV emission line profiles
of singly and doubly ionized species do not show the predicted trends of blue-shifted (out-flowing)
centroids \citep{harper01}. Remarkably, the low opacity line profiles of, e.g., \ion{C}{2}\up] 2325\,\AA\,{} and
Si~III\up] 1892\,\AA\,{}, in cool evolved stars tend to show a small red-shift, 
i.e., flows down towards the photosphere (\citealt{carpenter_etal91, carpenter_etal95, harper_etal95}).
For the particular case of the red supergiant Betelgeuse (M2~Iab, $\alpha$~Orionis, HD~39801)  
multi-wavelength Very Large Array (VLA) radio interferometry \citep{lim_etal98}
revealed that the atmosphere is cooler and significantly less ionized than the thermal structure 
predicted by the Alfv\'en wave-drive  model of \cite{hartmann_avrett84}.
[Note that while the dominant component is quite cool there is warm/hot material embedded within it
as indicated by  H$\alpha$ images, e.g., \cite{hebden_etal87}, 
and {\it HST} STIS spatially resolved chromospheric spectra of \ion{C}{2}\up] 2325\,\AA {} emission 
\citep{harper_brown06}.] 

The wind acceleration region, in the first few radii above the photosphere, 
is of prime interest for placing empirical constraints
on theories of mass-loss and is the focus of much research, e.g.,
\cite{crowley_etal08}, \cite{harper_etal05}, \cite{kirsch_etal01}, \cite{skinner_etal97}, and 
\cite{haas_etal95}.
This region is particularly important because it is where most of the energy is injected into the
wind and the mechanisms responsible are likely to be most manifest. The energy deposited
above and below the critical radius $(\sim 1.4 -1.9 R_\ast)$ controls the terminal wind speed and 
mass-loss
rates, respectively. {\it HST} has revealed that the UV emission line profiles used previously are not 
good diagnostics of the wind acceleration region but are instead 
revealing 
complex chromospheric geometries and flows of hot plasma. 
This is a result of
the exponential temperature sensitivity ($\propto n_e \exp\{-hc/\lambda kT\}/\sqrt{T}$) of the 
electron collisional excitation rates for UV emission and the exponential sensitivity of 
hydrogen ionization at
chromospheric temperatures. For example,  \cite{hartmann_avrett84} show for $\alpha$~Ori
$$
{n_e\over{n_H}} \sim A_{met} + \left[1 + {6.3\times 10^3 \exp\{1.18\times 10^5/T\}\over{\tau_{Ly\alpha}W(R/R_\ast)\sqrt{T}}}\right]^{-1},
$$
where $n_H$ is the total hydrogen density (H~I and H~II), $A_{met}$ is the abundance of metal ions, 
$\tau_{Ly\alpha}$ is the H~I Ly$\alpha$ optical depth and $W(R/R_\ast)$ is the geometric dilution factor. 
These factors allow the total UV flux from the star to be 
dominated by small volumes of high temperature plasma. 

To study the wind acceleration in outflows we therefore seek new emission line diagnostics 
that are less sensitive to the presence of hot chromospheric material. 
Such lines naturally occur at longer wavelengths, 
but unfortunately the stellar photospheric continuum rises 
strongly longward of the UV and swamps potential line emission. 
Beyond the photospheric flux peak, in the mid-IR (5-25$\mu$m), the photospheric continuum has
declined significantly and now, for M supergiants, the continuum becomes dominated by silicate
dust emission.
The mid-IR is also a good spectral region for optically thin emission 
line diagnostics. The longer wavelengths ensure much smaller Einstein decay rates, especially for
forbidden transitions, as compared to UV and optical emission lines and 
therefore mid-IR transitions are much less susceptible to
multiple scatterings in the wind that would make line profile interpretation 
more problematic. 
 We are interested in the tepid wind acceleration region so we also need to be able 
to distinguish its emission from the emission from the extended cold circumstellar envelopes (CSEs),
which are known to emit emission lines from ground terms of atoms and singly ionized species, 
i.e., \up[\ion{O}{1}\up] 63.18\,$\mu$m and
\up[\ion{Si}{2}\up] 34.81\,$\mu$m emission observed with the {\it Kuiper Airborne Observatory (KAO)}
\citep{haas_glassgold93} and the \up[\ion{Fe}{2}\up] 25.99\,$\mu$m and \up[\ion{Fe}{2}\up] 35.35\,$\mu$m 
observed with
{\it Infrared Space Observatory (ISO)} \citep{justtanont_etal99}. Suitable candidates for
wind acceleration diagnostics are emission lines from excited 
energy terms with $T_{exc} \sim 3000$~K since the excitation energy is well in excess of the available 
thermal energy in the CSE ($T_{gas} \simeq 100-1000$~K), and are also detectable from the ground.

In short, to study the wind acceleration in spatially unresolved spectra of M supergiants 
requires mid-IR diagnostics, a sensitive spectrograph with sufficient spectral resolution to resolve the line
profiles, and a telescope optimized for these wavelengths at a dry site: \up[\ion{Fe}{2}\up] $3d^7\>{\rm a\>^4F}$ emission, the
Texas Echelon Cross Echelle Spectrograph (TEXES) \citep{lacy_etal02}, and 
the infrared telescopes available 
on Mauna Kea is such a combination.

This paper can be considered as having two main parts. The first part is centered around
the TEXES observations of a sample of M supergiants and consists of 
\S2 which describes the new \up[\ion{Fe}{2}\up] diagnostics 
and their atomic data, \S3 which describes the TEXES observations of the M supergiants, and
\S4 which describes the empirical properties of the line profiles. 
The second part focuses on Betelgeuse for which different independent observations and 
atmospheric models are available to help interpret the new observations.
This part contains: \S5 which discusses the details of 
 \up[\ion{Fe}{2}\up] $3d^7\>{\rm a\>^4F}$ line formation as well as that for other 
well studied CSE emission lines;  \S6 which discusses the implications of our findings
for mass-loss mechanisms; and our conclusions which are presented in \S7.
Two appendices are included: the first describes the procedure to flux calibrate the 
TEXES spectra, and the second describes a composite model atmosphere for Betelgeuse 
that is used to calculate mid- and far-IR line fluxes.

\section{New Infrared Diagnostics}

\begin{figure}
\epsscale{0.8}
\plotone{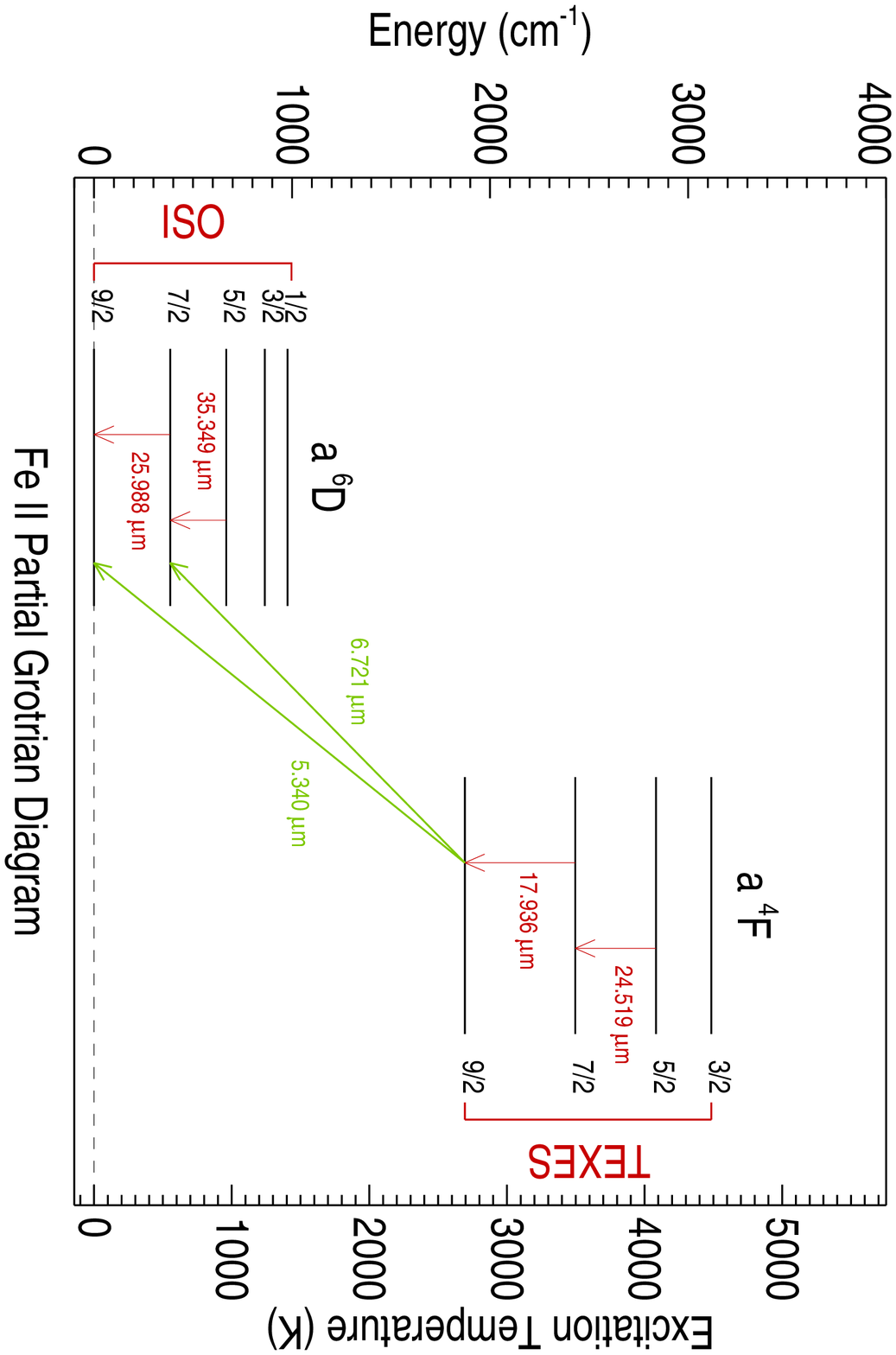}
\caption{Partial Grotrian diagram showing the
forbidden Fe~II lines observed with TEXES and ISO.\label{fig:grotrian}}
\end{figure}

Figure~\ref{fig:grotrian} presents a partial Grotrian diagram of the
two lowest energy terms of Fe~II showing the characteristic excitation temperature 
defined as (Energy$/k$).
The \up[\ion{Fe}{2}\up] 25.99\,$\mu$m and 35.35\,$\mu$m emission lines observed with {\it ISO}
are from within the ground $4s\>\>a\>^6D$ term ($0-977\>{\rm cm}^{-1}$) and probe the
cool CSE.  Here we use ``CSE lines'' to refer to emission from within ground energy terms, 
while the TEXES \up[\ion{Fe}{2}\up] lines have a hybrid character being from an excited term. 
For some emission lines this distinction is an oversimplifcation, e.g., for 
\up[\ion{Fe}{1}\up] 24.04\,$\mu$m where there may be a gradient in the ionization balance (\S5.3.1).
Observations of the \up[\ion{Fe}{2}\up] 25.99\,$\mu$m and 35.35\,$\mu$m kines 
were obtained with {\it ISO}-SWS 
at spectral resolutions of  $R\simeq 1000$ for $\alpha$~Ori, and $R\simeq 250$ for 
$\alpha$~Sco \citep{justtanont_etal99}; these resolving powers
are at least a factor of 20 too low to reveal either the 
turbulence or the flow dynamics. 
These transitions form a ladder which ends in the ground energy level 
($J=9/2$)
and can be used to constrain the wind temperature: 
35.35\,$\mu$m ($J_{ji}=5/2\to 7/2$),\footnote{We designate $j$ and $i$  
as the upper and lower levels of 
the emission lines, respectively.} 
and 25.99\,$\mu$m ($J_{ji}=7/2\to 9/2$). 

An analogous ladder exists within the next term $3d^7\>\>a\>^4F$ (i.e., the first excited term: 
$1872-3117\>{\rm cm}^{-1}$): 24.52\,$\mu$m ($J_{ji}=5/2\to 7/2$), and 17.94\,$\mu$m ($J_{ji}=7/2\to 9/2$).
Fig.~\ref{fig:grotrian} shows these ladder sequences.
These $3d^7\>\>a\>^4F$ transitions have been observed previously by \cite{kelly_lacy95} 
in $R\sim 10,000$ Irshell spectra \citep{lacy_etal89} of the $\alpha$~Sco (Antares)
system\footnote{A subsequent discussion of these $\alpha$~Sco observations: 
``Haas, Werner, \& Becklin (1996)'' was not published (M. Haas, priv. comm.)} and there
is also weak coincident emission in the {\it ISO} $\alpha$~Ori spectrum. Since the emission lines 
present in the
{\it ISO} spectra are unresolved, the emission line to continuum flux contrast will increase with increasing 
spectral 
resolution until the lines become resolved. The TEXES spectral resolution of $R\sim 50,000$ provides an 
opportunity to detect these
lines, resolve their line profiles at the $6\>{\rm km\>s}^{-1}$ level, 
and, with good signal-to-noise ratio,
determine the emission centroid velocities to $\sim 1\>{\rm km\>s}^{-1}$.

The 17.94\,$\mu$m line lies in a spectral region with water features, 
both telluric and photospheric.  A narrow telluric water line very close
to the [Fe~II] line can make the 17.94\,$\mu$m feature difficult to interpret,
depending on the Doppler shift.
In contrast, the 24.52\,$\mu$m line lies in a spectral region where the 
telluric attenuation varies very slowly across the line profile, 
making it more suitable for detailed emission profile analysis.
In Table~\ref{tab:atomic} we give the radiative atomic data for these diagnostics.

There is also potential emission from between the $a\>^4F$ and $a\>^6D$ terms, namely
6.72\,$\mu$m and 5.34\,$\mu$m which are also shown in Figure \ref{fig:grotrian}. 
A characteristic of the forbidden transitions in the lowest terms of Fe~II is that the radiative rates 
within a term 
are stronger than the rates between the terms \citep{nussbaumer_storey88}.
These lines are therefore expected to have weaker emission than the 17.94\,$\mu$m and 24.52\,$\mu$m 
lines and also to sit upon a brighter, more complicated, stellar continuum.

\subsection{Atomic Data}

\subsubsection{Radiative Data}

To utilize the high spectral resolution of the TEXES data 
requires accurate wavelengths, or wavenumbers, to establish the Doppler shifts of
the line emission.  
We have adopted the most accurate laboratory wavenumbers of
$407.8434 \pm 0.0009 \> (1\sigma) \>{\rm cm}^{-1}$ (24.52\,$\mu$m) and
$557.5364 \pm 0.0008 \> (1\sigma) \>{\rm cm}^{-1}$ (17.94\,$\mu$m)
which are from an ongoing project at Lund Observatory to improve atomic data for forbidden iron lines 
\citep{aldenius_johansson07}. The $1\sigma$ uncertainties
correspond to 0.66 and 0.43~${\rm km\>s}^{-1}$, respectively.

Accurate Einstein decay coefficients ($A_{ji}$) are also required if these 
lines are to provide thermodynamic constraints. 
\cite{garstang62} calculated the magnetic dipole and electric quadrupole transition probabilities 
for the 24.52\,$\mu$m and 17.94\,$\mu$m lines, with the magnetic dipole decay probabilties completely 
dominating. 
More recent computations by \cite{nussbaumer_storey88} and 
the IRON Project {\em SUPERSTRUCTURE} code presented by
\cite{quinet_etal96} are both in good agreement. The latter two sources give $A_{ji}$'s
that are the same, which we adopt here, and these in turn are the same as the \cite{garstang62} values 
at the precision of his Table~III.

\subsubsection{Collisional Data}

To establish whether the \ion{Fe}{2} energy levels of the emitting plasma are in Local
Thermal Equilibrium (LTE) or non-LTE, requires collision rates in and between the 
\ion{Fe}{2} $a\>^4F$ and $a\>^6D$ terms. \cite{pradhan_zhang93} presented 
electron collision rate coefficients for forbidden IR Fe~II transitions that have an 
estimated uncertainty of 10-30\%. Recently, however, \cite{ramsbottom_etal07} 
presented electron collision rates for temperatures that encompass those expected in M supergiant 
atmospheres that are lower by a factor of 2. These uncertainties are small in comparison
with estimates of hydrogen collision rates.

Detailed collision rates for neutral hydrogen collisions have not
been calculated, but estimates have been made for de-excitation rates that are of order 
$10^{-9}\>{\rm cm}^{3}{\rm s}^{-1}$
\citep{aannestad73,bahcall_wolf68}.  These are uncertain by an order of magnitude. If hydrogen
is partially ionized then the total collision rates will be dominated by
electron collisions, but in a cool photoionized stellar wind 
hydrogen and electron collision rates may be comparable. However,
if the gas has a sufficiently high hydrogen density then the
\ion{Fe}{2} level populations have a Boltzmann (LTE) distribution and 
the 24.52\,$\mu$m and 17.94\,$\mu$m  diagnostics will then be {\em insensitive} 
to the collision rates and {\em sensitive} to the accurately known Einstein A-values.  
Large mass column densities will also tend to inhibit photon losses and drive the level populations 
towards LTE.

In \S5.2.1 we find that in the line forming region the \ion{Fe}{2} $a\>^4F$ and $a\>^6D$ terms
are close to collisional equilibrium so that current uncertainties in theoretical
collisional excitation rates are of minor consequence to the interpretation of 
these mid-IR lines.

\section{TEXES OBSERVATIONS}

We have observed a sample of M supergiants, given in Table~\ref{tab:obslog},
with TEXES in high-resolution mode on the 3\,m IRTF 
on Mauna Kea. The data described here were mostly obtained in 2004 October, 
2005 January, and 2005 December (see Table~\ref{tab:obslog} for dates). 
 These observations are the longest wavelengths observed 
with TEXES, and were facilitated by a CdTe window that replaced the previous KBr window.
For the long wavelength observations we used a 2\arcsec\, slit width
(in the dispersion direction) to obtain the maximum spectral resolution
and typically nodded 6\arcsec\, along the 17\arcsec\, slit 
to subtract the sky emission. The detector pixels have a linear size of 0.33\arcsec,
providing a fully sampled Line Spread Function (LSF). 
The nodded star observations were interleaved with observations of a black 
thermal source and the sky. For our TEXES observations, very bright astrophysical 
sources with smooth continua suitable for
flat fielding, such as asteroids, were not often available. 
We therefore used the black thermal source and sky 
observations to provide an approximate flat field. We derived a first order correction for
telluric features, and an estimate of the sky and telescope transmission as described in \cite{lacy_etal02}.
The source flux ($F_\nu$), uncorrected for slit losses, is then given by

\begin{equation}
F_\nu \simeq S_\nu\left({\rm obj}-{\rm sky}\right) 
{B_\nu\left(T_{amb}\right)\over{S_\nu\left({\rm black}-{\rm sky}\right)}} 
\end{equation}
where $B_\nu\left(T_{amb}\right)$ is the Planck function for the
calibration source at temperature, $T_{amb}$, and $S_\nu$ is the
recorded signal.
 
The data were collected and initially examined at the telescope using a near real-time 
reduction package written 
in the {\it Interactive Data Language (IDL)} which facilitated
an efficient observing strategy. After the observing run the data
were carefully optimized during re-calibration with the pipeline reduction software. The spatial
profile on the detector, FWHM $\sim 2.5$\arcsec, was used to create a template
to extract the stellar spectrum.

We observed the wavelength regions covering the [Fe~II] lines discussed in \S2, as well as some of
the transitions between the $a\>^4F$ and $a\>^6D$ terms, and also the ground term [Fe~I] 24.04\,$\mu$m. 
Here we report on observations obtained so far; not all stars have been observed at all wavelengths. 
At long wavelengths, in high-resolution mode, only a portion of 4 spectral orders are recorded on the
detector at one time. 
The recorded regions are only about $200\>{\rm km\>s}^{-1}$ wide and do not overlap.  
Because the features we observe are
20\% of the order width, care must be taken to observe spectral 
features close to the center of the detector. 
An example of the spectral orders observed for the long wavelength lines in $\alpha$~Ori is
shown in Figure~\ref{fig:texes_format}.

\subsection{Wavelength Calibration}

The wavelength scale was established by identifying telluric molecular features
in adjacent orders. The wavelength calibration of the 24.52\,$\mu$m line, however, 
requires special mention. Finding suitable telluric features near this line proved impossible, 
so two different wavelength solutions were examined.
In the first case, the [Fe~II] line was observed in 2nd-order and then the
filter was changed to observe a telluric feature in 
the 12\,$\mu$m region in 4th-order. The grating equation was used
to establish the 2nd-order wavelength solution. Another solution 
was established using telluric features about 
8-orders from the emission feature. Both methods agree to 
$1 {\rm km\>s}^{-1}$ which is the level of desired accuracy for this science.

\subsection{Line Spread Function (LSF)}

The LSF was examined prior to mounting TEXES on the telescope using calibration water vapor 
spectra obtained from a low-pressure gas cell placed on the instrument entrance 
window. In the high resolution cross-dispersed operating mode
with a 2\arcsec\, slit (in the dispersion direction) the water lines near 24$\mu$m 
have emission cores that are well characterized by
a Gaussian with $R=\lambda/\Delta\lambda_{FWHM} \simeq 51,700 \pm 1600$.  At $20\mu$m the
resolution is $\sim 65,000$. The wings of the 
LSF are hard to quantify because the water lines sit upon a continuum.
In the following analysis we adopt a Gaussian LSF 
with $R=52,000\> [24/\lambda (\mu{\rm m})]$.

\subsection{Gemini-N Observations of $\alpha$~Scorpii (Antares)}

During the TEXES  Gemini-North engineering run in 2006 February,
a spectrum was obtained of the \up[\ion{Fe}{2}\up] 24.52 $\mu$m line in Antares on Feb 24. 
Unfortunately, no absolute wavelength calibration was obtained.
The slit was oriented north-south, perpendicular to the direction to the B 2.5~V star
companion. The slit was roughly 6\arcsec\ long and
we nodded the telescope east by 4\arcsec\ to remove sky emission and avoid potential
contributions from the companion that lies 3\arcsec\ west. 
The spectral resolution determined by the gas cell prior to the observing run was R=60,000 at
18.8\,$\mu$m, which suggests, via the scaling law used above, a spectral resolution at 24\,$\mu$m
of $6.4\>{\rm km\>s}^{-1}$.  

These data were reduced in the same manner as the IRTF data.

\section{RESULTS}

\subsection{Detections}

The stellar continuum near 17.94\,$\mu$m is quite structured in these evolved, oxygen-rich M
stars, and current limitations of theoretical M supergiant photospheric 
spectra preclude us from making positive detections unless the emission 
line stands above the adjacent continuum. 
In addition, the telluric interference near the 17.94\,$\mu$m line, depending on its Doppler shift, 
can make this line difficult to analyze.
We observed additional evolved K and M stars with widely different
surface gravities and mass-loss rates to provide an empirical check on the
structure of stellar photospheric continua.
A summary of the observations and
line detections is given in Table~\ref{tab:obslog}.
At 24.5\,$\mu$m the continuum
is less structured but TEXES is not as sensitive. However, the 24.5\,$\mu$m setting proved
useful in establishing the presence of \up[\ion{Fe}{2}\up] emission. We have dectected
\up[\ion{Fe}{2}\up] emission from all six of the M supergiants that we have observed.

Figure \ref{fig:texes_format} shows our first observations of $\alpha$~Ori at all three
wavelength settings. As mentioned before, we record only a portion of four orders at these
wavelengths.  The figure shows how the 17.94\,$\mu$m line might be
affected by a telluric water feature, while the sky transmission is
smooth for the other two lines. The \up[\ion{Fe}{2}\up] lines are much stronger than the
ground term \up[\ion{Fe}{1}\up] line which, when the emissivities are considered, 
shows that iron is predominantly singly ionized in the
extended atmosphere.

A comparison of $\alpha$~Ori with CE~Tau (M2~Iab) and $\mu$~Cep (M2~Ia) illustrates the effect 
of emission from circumstellar oxygen-rich dust \citep{sloan_price98} on the line-to-continuum 
ratio. Figure \ref{fig:cetau} 
shows the the \ion{Fe}{2} emission for CE~Tau which
is a close spectral-type proxy for $\alpha$~Ori and the line profiles are very similar. 
The line to continuum ratio is, however, substantially larger than observed in $\alpha$~Ori and
this is, at least in part, a result of the much weaker (or absent) dust emission from CE~Tau.  
$\mu$~Cep (shown in Figure \ref{fig:mucep}) has much stronger silicate dust emission 
and a greatly reduced line to continuum contrast. In $\mu$~Cep the 17.94\,$\mu$m line is also detected
but the two lines appear to have differently shaped profiles and the less symmetric
17.94\,$\mu$m line is slightly redshifted with respect to
the adopted $V_{rad}=+19.4\>{\rm km\>s}^{-1}$. This may be a result of underlying 
photospheric molecular features.

We also observed AD~Per (M2.5~Iab), our most distant source at $\sim 2$\,kpc, and detected the
17.94\,$\mu$m line which is shown in Figure \ref{fig:adper}. This star appears
to have unusual dust chemistry with carbon-rich dust (SiC) but an
oxygen-rich photosphere (\citealt{skinner_whitmore88,skinner_etal90}).

The difference between $\alpha^1$~Her (M5~II) and $\beta$~Peg (M2.5~II-III) 
in the 17.9\,$\mu$m region is shown in Figure \ref{fig:aher} which reveals a 
narrow emission component near the stellar rest frame of $\alpha^1$~Her. 
Subsequent 24.52\,$\mu$m observations, not described here, confirm the \up[\ion{Fe}{2}\up] emission. 
Both $\alpha$~Her's \up[\ion{Fe}{2}\up] profiles are slightly narrower than in the 
more luminous counterparts.
This figure illustrates the difficulty of identifying 
the emission line in this wavelength region, in the absence of reliable synthetic stellar spectra, 
when the emission is comparable in strength to the continuum.

The 24.52\,$\mu$m line of $\alpha$~Sco shown in Figure \ref{fig:antares} is 
slightly wider than for the other stars (see Table~\ref{tab:velocities}), however we find no indication of extended emission.
The Antares nebulae is a source of rich optical \ion{Fe}{2} emission, e.g., \citet{swings_preston78}, 
which is excited by the nearby B star companion (separation of 2.7\arcsec) 
\citep{reimers_etal08}. TEXES observations of this system may be sampling material from a slightly 
more extended, but still spatially unresolved,
region than for the single stars, and this material may have different velocity fields.

\subsection{Properties of Line Profiles}

To characterize the observed emission, Gaussian profiles have been fit to the  
spectra and their properties are given in Table~\ref{tab:velocities}. $\alpha$~Ori 
was observed on several occasions and we have also flux calibrated
these spectra as described in Appendix A. The fits to these individual spectra
are given in Table~\ref{tab:alphaori} which provides an indication of the
reproducibility of the spectra and of the intrinsic variability. 

The centroid emission velocities are given with respect to the adopted
stellar (center-of-mass) radial velocities which have typical uncertainties of
at least $1\>{\rm km\>s}^{-1}$. For M supergiants the center-of-mass radial velocities 
are not very well determined because
there are photospheric radial velocity variations, e.g., \cite{jones28}, typically with
amplitudes of $\pm 3-5\>{\rm km\>s}^{-1}$, that have both semi-regular 
(often with multiple periods of
hundreds of days to several years) and short-term erratic variations, 
e.g., \cite{smith_etal89}. So at any given time the photosphere
has a strong likelihood of moving with respect to the center-of-mass. Few M supergiants
have been monitored for sufficient lengths of time to determine 
$V_{rad}$ with sub-${\rm km\>s}^{-1}$ accuracy.

The observed emission centroid velocities are all close to the stellar center-of-mass 
radial velocities which suggests that the \up[\ion{Fe}{2}\up] emission is {\em not} 
formed in the convective churning photospheres which undergo velocity fluctuations seen
in optical absorption features. 
It is more likely
that the emitting region is larger and, or, decoupled from the variable surface layers. 
For $\alpha$~Ori we observe little, if any, variation with time 
which further supports this conclusion. During the 14 month period of our observations
$\alpha$~Ori's photospheric apparent velocity spanned a range of at least 
$3.7\>{\rm km\>s}^{-1}$ \citep{gray08}, although we do not have enough epochs
in common to study possible correlations.
It is well known that
the chromosphere and wind of $\alpha$~Ori is decoupled from its
photospheric variations \citep{goldberg79}. What is apparent is that
the \up[\ion{Fe}{2}\up] emission profiles are neither blue-shifted nor are they flat-topped. Both
of these properties exclude the possibility that the emission is formed with any significant
outflow velocity, typically $\sim 10\>{\rm km\>s}^{-1}$ for M supergiants. 
If the emission were from within such a moving flow, the absence of blue shifted profiles excludes the flux 
being formed close to the star, while the absence of a top-hat profile excludes the 
emission originating from an extended region. We conclude the emission arises from  
material with at most a small outflow velocity. Further consideration of the 
line formation in \S5.1 reveals that the emission arises from close to the star.

The observed line profiles are well resolved.  Because the cores of the TEXES LSF and 
observed profiles are quite Gaussian, the intrinsic stellar {\it most probable} 
turbulent velocity ($V_{turb}$), which we assume to be 
isotropic\footnote{FWHM$=2\sqrt{\ln{2}}V_{turb} =1.6651 V_{turb}$.} , can be estimated from
\begin{equation}
V_{turb} = \sqrt{V_{Dopp}^2{\rm (Obs)} - (X\>{\rm km\>s}^{-1})^2},
\end{equation}
where $X=2.6$ and $3.5\>{\rm km\>s}^{-1}$ are the corrections for the TEXES instrumental broadening
at 17.94 and 24.52\,$\mu$m, respectively. The observed \up[\ion{Fe}{2}\up] line widths given in 
Tables~\ref{tab:velocities} and \ref{tab:alphaori} are similar with a range of
$V_{turb}=12-15\>{\rm km\>s}^{-1}$ with $\alpha$~Sco  having the largest value.

For $\alpha$~Ori, the \up[\ion{Fe}{2}\up] line widths are similar for both the 17.94\,$\mu$m and 
24.52\,$\mu$m lines ($\simeq 12.5$\kms) and 
they do not change significantly between the three different observing runs. 
The 24.04\,$\mu$m \up[\ion{Fe}{1}\up] line is significantly narrower and
indicates a different line forming region. Both of these forbidden line widths are significantly less than 
that derived from the chromospheric UV C~II\up] 2325\,\AA {} emission multiplet. 
The {\em sky integrated} C~II\up] profiles
observed with {\it HST} have non-Gaussian profiles whose FWHM implies $V_{turb}=19-21$\kms
\citep{carpenter_robinson97}.
Radiative transfer modeling of the {\em spatially resolved} 
{\it HST}/STIS C~II\up] 2325\,\AA {} emission reveals that 
these lines are slightly opacity broadened at the stellar limb and can be well matched with 
intrinsic turbulence of 
$V_{turb}=17-19$\kms which changes slowly over large spatial scales: 
$1.5R_\ast < R < 3.5 R_\ast$ \citep{harper_brown06}. This is the same spatial region over which we 
anticipate that the \up[\ion{Fe}{2}\up] emission originates.

The cool component of $\alpha$~Ori's inhomogeneous atmosphere, traced by thermal radio 
continuum observations, has now been dynamically resolved from the hot component,
traced by UV emission lines, for the first time.  The
\up[\ion{Fe}{2}\up] profiles, with their much lower temperature sensitivity, reflect the amplitude of the 
motions in the cooler plasma which are less than that of the hotter chromosphere. Since the cool atmospheric component includes the base of the
wind outflow, it is these lower amplitude motions that should be associated with the
unknown wind driving processes. For 1000-3500~K plasma these turbulent velocities, if interpreted as occurring on 
small spatial scales, imply significant Mach numbers. While the TEXES \up[\ion{Fe}{2}\up] profiles are 
spatially unresolved (they are global averages), there is no evidence for outward travelling shocks moving with 
these velocities in the line forming region.

For our TEXES \up[\ion{Fe}{2}\up] detections the turbulent velocities are similar in all stars, 
which may not be a surprise since the sample consists of mostly early M supergiants. The remarkable discovery 
by \cite{lim_etal98} 
that the extended atmosphere of $\alpha$~Ori is dominated by cool, rather than hot gas as previously thought, 
has now been confirmed for $\alpha$~Sco with VLA A-configuration observations made by Brown \& Harper \citep{harper09}. 
The presence of
extended cool non-chromospheric plasma with $V_{turb}\simeq 13$\kms 
is likely a common property of early M supergiants and not a rare curiosity, and
deserves further attention.

These \up[\ion{Fe}{2}\up] turbulent velocities are larger than the macroturbulence required to model 
upper photospheric 12\,$\mu$m molecular OH and H$_2$O absorption lines of 
$\mu$~Cep \citep{ryde_etal06b} and 
$\alpha$~Ori \citep{ryde_etal06a}. 
The 8\kms turbulence\footnote{The most probable micro- and macro turbulence velocities added in quadrature.} required to
match $\alpha$~Ori's  12\,$\mu$m TEXES spectrum in \cite{ryde_etal06a}
is actually smaller than that needed to
model the optical: 11\kms (\citealt{gray00,gray08}) and $\sim 15$\kms \citealt{gray01}), and
near-IR: 12\kms  \citep{lobel_dupree00} photospheric lines. The conclusion to be drawn  from
this is that as absorption lines are formed farther out
from the star, they become less sensitive to the vigorous photospheric convective motions, which in turn
is reflected in the lower macroturbulence required to match the observed line widths. At some
radius where the extended atmosphere becomes decoupled from the photosphere, 
the turbulent motions increase once more in both the hot chromospheric and cool
wind components.

\subsection{Thermal Constraints}

To place the dynamical information from the resolved line profiles in better
context we need to establish where the emission is formed. In this subsection we will
consider the most general formation properties and then in \S5 we will consider the
contribution functions of the TEXES and {\it ISO} CSE lines from $\alpha$~Ori in more detail.

The characteristic formation temperature can be derived by assuming
that the relative level populations of the upper (j) and lower (i) energy levels $(n_j, n_i)$ 
can be described by a Boltzmann distribution with a characteristic excitation temperature
$(T_{exc})$ where
\begin{equation}
{n_j\over{n_i}} = {g_j\over{g_i}}\exp\left\{-(E_j-E_i)/kT_{exc}\right\}.
\end{equation}
The $g$'s are the statistical weights, and $E_j-E_i$ is the energy
difference between the upper and lower energy levels. If the wind is isothermal and the energy levels are in 
thermal equilibrium then $T_{exc}=T_{gas}$. From the Einstein A-values and by assuming optically thin 
emission for the ratio of the {\it ISO} fluxes, \cite{justtanont_etal99} derive excitation temperatures 
for the ground term emission of
$T_{exc} \simeq 1785$\,K and $T_{exc}\simeq 1230$\,K for $\alpha$~Sco and $\alpha$~Ori, respectively. 
From the ratio of populations in the ground $a\>^6D$ and excited $a\>^4F$ terms the 
TEXES $\alpha$~Ori \ion{Fe}{2} fluxes give $T_{exc} \sim 1520-1950$\,K. Since in $\alpha$~Ori 
the ground state emits at a lower characteristic temperature, this provides
a lower limit for $T_{exc}$ for the  $a\>^4F$ excitation region. For the ratio of fluxes 
within the  $a\>^4F$ term we find a $3\sigma$ lower limit of $2110$\,K, so 
the atmosphere is not isothermal. 

Figure \ref{fig:te} (in Appendix B) shows the composite temperature structure for $\alpha$~Ori
described in Appendix B and also shows the formation radii based on the 
{\it ISO} and TEXES temperature constraints under these simple isothermal and optically thin
assumptions. In the theoretical model of \cite{rodgers_glassgold91}
the $a\>^6D$ ground term emission originates near $\sim 10R_\ast$, however,
we now know from the thermal radio continuum observations of \citet{lim_etal98} 
that this temperature must occur slightly closer to the star, i.e., at $\sim 7.4R_\ast$.
The $a\>^4F$ emission originates interior to this at $\le 4.1 R_\ast$ where the
outflow velocities are expected to be small. This spatial constraint provides a 
partial explanation of why Doppler blue-shifted wind signatures are not observed. 

In summary, we find that the M supergiants share common properties in that their 
mid-IR \up[\ion{Fe}{2}\up] line profiles
appear to be quite Gaussian (rather than top-hat) and show no evidence of significant
Doppler shifts indicative of outflow. To within the combined uncertainties the lines are 
at rest in the stellar rest frame. The line-to-continuum contrast is a function of the 
circumstellar dust emission as expected. The characteristic excitation temperature 
places the line formation close to the star where the outflow velocities are expected
to be low, which explains the lack of a clear wind signature. For the
case of $\alpha$~Ori the line profiles do not show significant variability 
and the line widths are systematically smaller than those observed in spatially 
resolved UV spectra of the hotter chromosphere. 
We have dynamically resolved the turbulent motions in the
dominant and pervasive cool atmospheric component.

\section{DISCUSSION: $\alpha$~Orionis in Context}

{\em Where are the mid- and far-IR emission lines observed in $\alpha$~Ori with TEXES, {\it KAO}, 
and {\it ISO} formed?}
To quantify the emission contributions from 
different radii, a thermodynamic and dynamic model is required that encompasses the 
chromosphere, inner wind, and CSE.
Currently no such comprehensive models exist. $\alpha$~Orionis provides the best studied 
example of an M 
supergiant and we will use the 
properties of this star throughout this section to quantify the mid- and far-IR
line emission, with the reasonable assumption that the results will 
apply at some level to early M supergiants in general. 

While no complete atmospheric model exists, models do exist for the inner region
\citep[HBL01 hereafter]{harper_etal01} and the outer 
CSE \citep[RG91 hereafter]{rodgers_glassgold91}.
Appendix B describes a spherical (1-D) composite dynamic ($V_{turb}(R)$, $V_{wind}(R)$) and 
thermodynamic ($T_{gas}(R)$, $\rho(R)$) model that utilized these earlier results
and interpolates between them. This composite model is essentially a 
combination of the spatially extended semi-empirical model of
HBL01 scaled to the recently revised stellar 
distance \citep{harper_etal08} and one of the variational thermodynamic 
models of RG91, and 
is referred to as the {\it Composite Model Atmosphere}. In the following calculations we use the
cooler inner wind model which is shown as a dashed line in Fig.~\ref{fig:te} in Appendix B.

The HBL01 inner region model was based upon multiwavelength spatially-resolved VLA data 
covering 0.7-6~cm
combined with non-contemporaneous spatially unresolved data at shorter wavelengths. 
The HBL01 model predicts a thermal continuum flux at 100~GHz (0.3\,cm) of 92.2~mJy which is
insensitive to the wind dust emission.  
As part of a larger multiwavelength study of M supergiants 
and to provide a check on temporal changes in the 
extended atmosphere of $\alpha$~Ori, we obtained observations 
of $\alpha$ Ori, $\alpha$~Sco and $\alpha$~Her at 100~GHz with 
the OVRO\footnote{The Owens Valley Radio Observatory was supported by
the National Science Foundation, AST 9981546.} Millimeter Array and these
are described next.

\subsubsection{Owens Valley Radio Observatory (OVRO)}

The OVRO observations of $\alpha$~Ori, $\alpha$~Sco, and $\alpha$~Her
are summarized in Table~\ref{tab:ovro}. For $\alpha$~Ori, observed on
2003 November 9, four 1 GHz continuum 
bands were observed with the dual-channel analog correlator centered 
around 100 GHz and spanning
the range 96.5-103.5 GHz. The antennae were in the L configuration with
baselines between 15-115 m, although only 5 antennas were available
during the observation. The instrumental gain was calibrated every 15
minutes using the quasar J0532+075. The absolute flux was bootstrapped
from J0923+392 observations, because no planets were available during
the $\alpha$ Ori transits, resulting in a 15\% uncertainty in the
absolute flux scale.  The calibrations were done with the OVRO MMA software
\citep{Scoville93} and the images were produced using standard
routines in Miriad \citep{sault95}.

For $\alpha$ Sco and $\alpha$ Her, observed in a shared track on 2004
March 30, the gains were calibrated using the quasars J1517-243 and
J1608+104, respectively and fluxes were bootstrapped from these two
quasars with a similar 15\% uncertainty. The correlator setup was the same as for $\alpha$~Ori. 
This track was taken in E configuration which contains several more extended
baselines than L with baselines between 35 and 119 m and all 6
antennae were present throughout the track.

It is interesting to compare the 100~GHz fluxes with the 250~GHz fluxes
measured by \cite{altenhoff_etal94} and shown in Table \ref{tab:ovro}.
At these high frequencies the earlier spectral-type companions of 
$\alpha$~Sco and $\alpha$~Her should have
negligible flux contributions, e.g. \citet{hjellming_newell83}. The 250/100~GHz flux ratios for
$\alpha$~Sco, $\alpha$~Ori, and $\alpha$~Her are $4.9\pm 0.5$, $4.4\pm 0.4$, and 
$5.5\pm 0.5$, respectively. When the 100~GHz fluxes are normalized to the product of the
star's effective temperature and angular diameter squared, e.g., from \cite{dyck_etal96},
the two luminosity class Iab M supergiants ($\alpha$~Ori and $\alpha$~Sco) have 
similar ratios (within 10\%) while for the less luminous 
$\alpha$~Her, the ratio is about half this value, which may reflect it's less massive
extended atmosphere.

The HBL01 model predicted a 100~GHz flux (92~mJy) which is consistent with the rather uncertain 90~GHz fluxes recorded 
in 1975 (\citealt{newell_hjellming82} and 
references therein). Our OVRO $\alpha$~Ori flux was recorded about a year
before the first TEXES observations and is slightly lower which 
may reflect the mean atmospheric temperature being slightly cooler than 
adopted in HBL01.

\subsection{Line Formation} 

Here we examine the line formation of the forbidden excited \up[\ion{Fe}{2}\up] and ground term 
CSE lines by computing their emission profiles and contribution functions. 
We assume that the  source functions of the 
relevant forbidden lines 
are in Local Thermal Equilibrium (LTE), i.e., $S_\nu \simeq B_\nu\left(T\right)$.
This can be a reasonable approximation when the particle densities 
in the line formation region are greater than the critical 
densities. These conditions can be checked {\it a posteriori} and are discussed 
further in \S5.2.1.

We include the wind and turbulent velocity fields in the atomic absorption profile (which is 
assumed equal to the emission profile, i.e., {\it complete redistribution}) 
and compute the resulting spectral profiles from the formal
solution of the equation of radiative transfer in a spherical atmosphere
which runs from the upper photosphere through to the CSE.

The adopted abundances and Einstein decay coefficients are given in Tables~\ref{tab:atomic} and 
\ref{tab:cse_lines}.  For the thermal conditions in the extended envelope, stimulated emission is
important for these mid- and far-IR transitions.
The background continuous opacity is 
dominated by pure absorption and has contributions from bound-free opacity
from excited levels of neutral species, and H and H$^-$ free-free opacity. 
At these wavelengths, the bound-free opacity comes 
from hydrogenic quantum numbers $n\ge 14$, which are likely to be partially 
collisionally coupled to the continuum, so that both bound-free and H free-free opacity 
are proportional to the electron density. We performed trial non-LTE calculations for the
populations of the hydrogenic Rydberg $(n,l)$ states, following \cite{hummer_storey92}, and
found that, at the large column densities where the continuum opacity is important 
for these mid- far-IR lines, the departure coefficients of the $(n,l)$ levels are not 
significantly different from unity. The continuous opacity is important in
the deeper layers because of the density sensitivity, $\kappa_{cont} \propto n_e n_H$, and 
the continuum sets the inner boundary condition for the line
formation problem. The modelled line fluxes were obtained by measuring the emission above the 
computed local continuum and are given in Tables~\ref{tab:feii_observed} and
\ref{tab:fluxes_other}.

\subsection{Flux Contribution Functions}

An alternate way to estimate the line fluxes is to sum up the emission from each
volume element in the extended atmosphere and wind, i.e.,
\begin{equation}
F_\oplus \simeq {h\nu_{ji} A_{ji}\over{D^2}}
\int\limits_{R_\ast}^\infty n_H {n_{FeII}\over{n_H}}{n_j\over{n_{FeII}}} P_{esc}(R) R^2 \> dR
= {h\nu_{ji} A_{ji}\over{D^2}}
\int\limits_{R_\ast}^\infty n_H {n_{FeII}\over{n_H}}{n_j\over{n_{FeII}}} P_{esc}(R) R^3 \> d\ln{R}
\end{equation}
where $D$ is the distance to the star, $h\nu_{ji}$ is the photon energy, 
$A_{ji}$ 
is the Einstein decay coefficient, $n_H$ is the total hydrogen population,
$n_{FeII}/n_H$ is the abundance of
\ion{Fe}{2} relative to hydrogen ($=A_{FeII}$),  $n_j/n_{FeII}$ is the ratio 
of the
population of the upper emitting level to the total \ion{Fe}{2} population, 
and $P_{esc}(R)$ is the single-flight escape probability
of the photon emitted at radius $R$. This escape probability is 
adopted because the particle densities are high enough that the
probability of a photon scattering and subsequently escaping is small. 
The high particle densities indicate that 
fine-structure level populations will be close to a Boltzmann distribution, so that 
$$
{n_j\over{n_{FeII}}} \simeq {g_j\over{U(T_{gas})}}\exp\left\{-E_j/kT_{gas}\right\}
$$
where $U(T_{gas})$ is the non-LTE partition function of \ion{Fe}{2}, 
and $E_j$ is the energy of the
emitting (upper) level with respect to the ground energy level. 

The inner boundary is chosen to be deep enough that
$P_{esc}\left(R\right)\to 0$. The escape probability takes into account 
photons that are thermalized during line scattering and by continuum absorption. 
For low optical depths at a few stellar radii $P_{esc}$ is approximately
the fraction of the sky not subtended by the star and rapidly approaches
unity as the radius increases. Closer to the star $P_{esc}$ allows for
photons that escape in the wings of the emission line when the 
line center optical depth is greater than unity.

To illustrate where the emission lines originate we define a radially weighted 
contribution function
\begin{equation}
C_{rw} = n_H {n_{FeII}\over{n_H}}{n_j\over{n_{FeII}}} P_{esc}(R) R^3 
\label{eq:crw}
\end{equation}
such that, when plotted against $\ln{R}$, the area under the curve
shows the relative contribution of different regions to the total flux.
While the formal solution of the transfer equation provides, in principle, 
an {\em exact} $P_{esc}(R)$,
here we use Eq.~(\ref{eq:crw}) to illustrate $C_{rw}$.

Convenient expressions for the escape probability have been derived for 
plane parallel geometry and certain spherical distributions of static 
scattering material \citep{kunasz_hummer74}. 
For stellar winds where  $V_{wind} > V_{turb}$ the Sobolev escape probability is
often employed \citep{castor70}. Normalizing radial distances by the stellar radius, i.e., $Z=R/R_\ast$,
the continuum radial optical depth of 
unity occurs close to the stellar surface at $Z_{cont}$. Here we approximate 
$P_{esc}(R)$ as the larger of the Sobolev value or
\begin{equation}
P_{esc}\left(R\right) = K_2\left(\tau\right) E_2\left(\tau_{cont}\right){1\over{2}}
\left[1+\sqrt{1-\left({Z_{cont}\over{Z}}\right)^2}\right]
\end{equation}
where $K_2$ is the half-sky plane-parallel Doppler profile single-flight
escape probability given by \cite{hummer81} 
with the mean optical depth of a static atmosphere\footnote{The argument of 
the kernel $K_2$ is the mean optical depth, and for a Doppler
profile $\tau=\tau_0 \sqrt{\pi}$ where $\tau_0$ is the static line center
optical depth.}. The $E_2$ term approximates the fraction of line photons not lost to the continuum 
and is only important close to the star. The geometric term allows for stellar 
occultation, and the factor $Z_{cont}\sim 1.2$ is the radius 
where the tangential continuum optical depth is unity. 

Figure \ref{fig:contrb} shows the normalized $C_{rw}$ for the TEXES and CSE lines.  
The narrow peak at $Z\simeq 1.5$ corresponds to the high particle densities and
maximum temperatures in the {\em Composite Model} with the sharp cut-off on the
photospheric side resulting from continuum absorption which affects all lines
in a similar fashion and eliminates any photospheric contributions to the emission fluxes. 
The decline outward of the peak is a combined result of
the declining $T_{gas}$ and density. Previous generations of theoretical, e.g.,
\cite{hartmann_avrett84}, and semi-empirical models, 
e.g., \cite{wischnewski_wendker81} and \cite{lobel_dupree00},
had more extended warm chromospheres. 
Most of these can be ruled out by the observed
narrowness, the absence of observed blue-shifts or wind broadening  
in the TEXES \up[\ion{Fe}{2}\up] profiles. 
The small discontinuity in the figure at $Z\simeq 7$ reflects where the
density structure of the outer wind has been merged with the inner density
structure that is constrained by radio observations (see Appendix B).

The TEXES and {\it ISO} \up[\ion{Fe}{2}\up] lines clearly have 
different formation radii as previously suggested by their different 
characteristic excitation temperatures.
The TEXES lines have half their emission from a region around the peak-$T_{gas}$ ($Z=1.5$) 
while the {\it ISO} lines are formed around $Z\sim 6$. The absence
of wind shifted emission in the TEXES lines is a result of
the significant contribution from the quasi-static region at chromospheric radii
and above. It is thought that this extended region, resolved with the VLA by
\cite{lim_etal98}, is in the base of the wind where the velocity is 
small \citep{harper_etal01} and thus the TEXES \up[\ion{Fe}{2}\up] lines are probing the wind, albeit 
at low outflow velocities. 

Fig.~\ref{fig:contrb} suggests that the wind acceleration signature 
might be apparent in the ground term \up[\ion{Si}{2}\up] 34.81\,$\mu$m and 
\up[\ion{Fe}{2}\up] 35.34 and 25.98\,$\mu$m profiles if observed with sufficient spectral resolution.
Note that these lines have a non-negligible flux contribution from within
$3R_\ast$ which, however, is often taken as the {\em inner boundary condition}
(\citealt{rodgers_glassgold91,haas_glassgold93,haas_etal95,justtanont_etal99}).
Clearly consideration of the material at $R < 3R_\ast$ is required for detailed
analysis of these mid- far-IR lines. 

The \up[\ion{O}{1}\up], \up[\ion{C}{1}\up], and \up[\ion{C}{2}\up] lines are
expected to have top-hat profiles and be centered close to the stellar rest frame, 
as the fraction of red-shifted emission 
occulted by the star is tiny. Indeed observation of \up[\ion{C}{1}\up] 609\,$\mu$m
by \cite{huggins_etal94} reveal that this line has similarities to
mm-CO emission profiles and suggests that its very broad spatial contribution function
also includes material traveling with the faster S2 shell (see Appendix \S B2.1). 

The \up[\ion{Fe}{2}\up] emission sits upon upper photospheric/lower chromospheric 
absorption and more reliable model fluxes require a more detailed description 
of the $T_{gas}$ structure between the 
chromospheric $T_{gas}$-rise and the photosphere than available at present. 
Recent VLTI MIDI 7.5-13.5\,$\mu$m observations \citep{perrin_etal07} suggest the presence of 
a cool molecular rich region with $T_{gas} \sim 1550$\,K interior to 1.25$R_\ast$ (corrected to 
the angular diameter adopted here). The presence of molecular material between the
upper-photosphere and the chromospheric temperature peak is reminiscent of the
bifurcated outer atmosphere observed off the solar limb in CO 
\citep{ayres02} albeit on a larger fractional radial
scale as befits the lower surface gravity of Betelgeuse. This molecular material 
may also be related to the detection of water vapor in the outer photosphere of Arcturus
by \cite{ryde_etal02}. 
Future {\it Atacama Large Millimeter Array (ALMA)} interferometric sub-mm continuum observations will provide
independent thermodynamic constraints down from the chromosphere towards the photosphere 
and covering this intriguing molecular region. Such ALMA data will complement those 
from the VLA that sample the chromosphere and wind.

\subsubsection{Line Source Functions}

Departures from the assumed LTE line source functions can lead to uncertainties in the contribution 
functions and line fluxes. Potential departures can be considered by 
examining the {\em equivalent two-level atom} description of the line source function $S_L$ \citep{mihalas78}
\begin{equation}
S_L = {\bar{J} + \left(\epsilon^\prime +\theta\right) B_\nu\left(T_{gas}\right)
\over{1 + \epsilon^\prime + \eta}}
\label{eq:etla}
\end{equation}
where
\begin{equation}
\epsilon^\prime \equiv C_{ji}\left(1 - e^{-h\nu/kT}\right)/A_{ji}
\end{equation}
and $C_{ji}$ is the collisional de-excitation rate. $B_\nu(T_{gas})$ is the Planck function at
the line frequency, $\bar{J}$ is the mean intensity averaged over the line profile, 
and the other terms $\theta$ and $\eta$ represent the radiative and collisional coupling 
between the 
levels $i,j$ and all other energy levels of the ion, i.e., many possible interactions 
(see \citealt{mihalas78} for details).

If the net rate of radiative and collisional coupling between the upper (j) [and lower (i)] 
level of the mid- and far-IR transition to all other levels, excluding i (or j), can be neglected, 
(i.e., $\theta \ll \epsilon^\prime$ and $\eta \ll \epsilon^\prime$) 
then Eq. (\ref{eq:etla}) reduces to the standard two-level atom description.
Under these circumstances, if the downward collision rate for $j\to i$ 
is higher than the radiative decay rate ($A_{ji}$), 
then $\epsilon^\prime\gg 1$ and the levels are in an LTE ratio with $S_L \simeq B(T_{gas})$.
Estimates of the critical particle densities required to establish thermal equilibrium 
between the energy levels of forbidden lines are given in the compilation 
of \cite{hollenbach_mckee89}.
Although the hydrogen collision rates are very uncertain, typically
thermalization requires $n_H > 10^4\>{\rm cm}^{-3}$, which is satisfied 
for radii $ < 100R_\ast$. This is a necessary, but not sufficient, condition for
 $S_L \simeq B(T_{gas})$ because the coupling between other energy levels embodied by 
$\theta$ and $\eta$ can be important. 

For \ion{Fe}{2}, collisions play a particularly important role because the first 
64 fine-structure energy levels have the same parity and hence are coupled by collisions
that compete with parity conserving electric quadrapole (E2) and magnetic dipole (M1) 
transitions. Because there are so many energy levels the 
terms $\theta$ and $\eta$ are not simply evaluated, so to check the accuracy 
of the LTE source function approximation for the TEXES \up[\ion{Fe}{2}\up]
lines we have examined their source functions. Escape probabilities were used to
approximate the net radiative brackets in an \ion{Fe}{2} model with the first 769 energy levels
for the HBL01 model of $\alpha$~Ori. The atomic data are essentially those described
by \cite{sigut_pradhan98}. The ratio $S_L/B(T_{gas}) \simeq 1$ for $R < 7R_\ast$ with departures
of at least 10\% occuring in the outer line forming region

Photoexcitation by chromospheric ultraviolet radiation in the allowed transition 
\ion{Fe}{2} multiplets whose lower term is also $a\>^4F$,
e.g., Multiplet Nos. 20-31 $(A_{ji} \sim 10^5 - 10^{7}\>{\rm s}^{-1})$ \citep{fuhr_wiese06}
can lead to excitation depopulation rates in excess of the \up[\ion{Fe}{2}\up] decay rates.
These transitions are opaque and the depopulation rates depend on self-shielding which is 
sensitive to the 
wind velocity and turbulent gradients. Evaluating these rate is beyond the scope of the 
present work but we note that detailed non-LTE source functions are desirable for 
future analysis.

\subsection{Observed and Predicted Mid- and Far-IR Fluxes}

A comparison of the computed and observed fluxes for $\alpha$~Ori 
in Tables~\ref{tab:feii_observed} \& \ref{tab:fluxes_other} reveals a rather unusual
mismatch that is a function of formation radius. The computed TEXES \up[\ion{Fe}{2}\up]
fluxes are $\sim 3.1$ too large, the \up[\ion{Si}{2}\up] and {\it ISO} \up[\ion{Fe}{2}\up]
emission lines are $\sim 1.6$ too large, while the
lines formed at larger radii are in reasonable agreement with, or slightly underestimate,
the more uncertain observations.

There are different uncertainties in the calibrations and flux measurements of these lines
(observed with TEXES, {\it ISO}, and {\it KAO}) that arise from different elements with their 
inherent uncertainties in abundance and ionization state. However, because of the overlapping formation radii this 
systematic trend is hard to explain in a simple way. 
These mid- and far-IR lines have large contributions inside the silicate dust shell 
observed at $\sim 30R_\ast$ \citep{danchi_etal94}
and molecular abundances and the dust/gas mass ratio are lower than for cooler M supergiants, 
suggesting that the CSE flux discrepancy is not a result of depletion from dust formation or 
molecular chemistry. The combined uncertainties resulting from the observed fluxes and
intrinsic variability should be $< 30$\%, so next we explore other possible explanations.

\subsubsection{Ionization Balance?}

\subsubsubsection{Iron}

The \up[\ion{Fe}{1}\up] 24.04\,$\mu$m emission arises from the ground term 
and, for a {\em fixed ionization balance}, it might be expected to be formed 
in the same region as the {\it ISO} \up[\ion{Fe}{2}\up] 25.99\,$\mu$m line. The observed ratio of TEXES 
\up[\ion{Fe}{2}\up] to \up[\ion{Fe}{1}\up] fluxes in $\alpha$~Ori shows
that iron is predominantly singly ionized, in agreement with theoretical
calculations of \cite{rodgers90}. This allows the fluxes of the \up[\ion{Fe}{2}\up]
lines to be used as diagnostics of the amount of material in the extended atmosphere.
The \up[\ion{Fe}{1}\up] 24.04\,$\mu$m flux can be reproduced with the {\em Composite Model
Atmosphere} by assuming a constant $A_{Fe I}= 10^{-2}A_{Fe II}$.
However, the narrowness of the 
profile suggests that it has a stronger contribution from closer to the star 
where the turbulence is smaller and thus the ionization of iron increases with radius above the surface.
The ionization balance in the chromosphere and inner wind region is controlled by the competing forces
of photoionization by the strong stellar UV radiation field and radiative recombination.
The \up[\ion{Fe}{1}\up] 24.04\,$\mu$m profile is less Gaussian and slightly asymmetric as 
compared to the TEXES \up[\ion{Fe}{2}\up] lines and possibly has a small blue shift,
in which case the \up[\ion{Fe}{1}\up] may have a wind emission component.

\cite{castro_carrizo_etal01} have reported a \up[\ion{Fe}{1}\up] 24.04\,$\mu$m 
flux from {\it ISO} grating
spectra of $3.5\pm 0.4\times 10^{-19}\>{\rm W\>cm}^{-2}$ which 
is significantly larger than we estimate from our TEXES spectra 
($7.7\pm \times 10^{-20}\>{\rm W\>cm}^{-2}$) which has a factor
40 greater spectral resolution. Fig.~\ref{fig:texes_format} 
reveals that there is another emission feature nearby which would be 
unresolved in {\it ISO} spectra
might account, in part, for the difference in measured flux. 
We are unable to identify this feature, but it is redshifted roughly 28\kms 
from the peak of the \up[\ion{Fe}{1}\up] emission. Therefore, we believe it is 
unlikely to be a separate component of \up[\ion{Fe}{1}\up].
\cite{aoki_etal98} have also reported the 
detection of this \up[\ion{Fe}{1}\up] line
in {\it ISO} spectra in two carbon stars (TX~PSc and WZ~Cas), but it was not
observed in the oxygen-rich giant 30~Her (M6~III).  These {\it ISO} observations 
suggest that even in this late-M giant there is sufficient UV flux to
photoionize low ( $< 13.6$~eV) ionization potential metals, while in 
the carbon stars the iron is less ionized.

The predicted flux of the {\it ISO} \up[\ion{Fe}{1}\up] 
$34.71 \mu$m is consistent with the observed upper-limit from 
\cite{castro_carrizo_etal01}.

\subsubsubsection{Other Elements}

Silicon is expected to be photoionized by the stellar UV radiation field and  
predominanty in \ion{Si}{2}, while 
\ion{O}{1} is expected to be the dominant ionization state. For carbon the ionization
balance is more uncertain. \ion{C}{2} dominates in the outer reaches of the CSE 
as the Galactic radiation field
ionizes any remaining \ion{C}{1} \citep{mamon_etal88}. Uncertainties in the ionization states 
do not appear to be the cause of the systematic discrepancies between the model and mid- and far-IR 
fluxes.

\subsubsection{Temporal Variability?}

The fluxes given in Table~\ref{tab:fluxes_other}  were observed over many years, and although there
are hints of intrinsic variability these are at the same level as the uncertainties in the flux
measurements. In some cases there may be off-source emission in the observing apertures which may 
mimic stellar variability \citep{haas_etal95}.  The TEXES observations do not indicate 
significant short time variations, and the {\it ISO} fluxes
were obtained shortly after the VLA observations used to construct the inner 
part of the atmospheric model. Therefore, we do not expect temporal variations sufficiently large 
to explain the
model/observed flux disagreement.

\subsubsection{Temperature and Density Distribution?}

When $T_{gas} \ge T_{exc}$ the line emissivity is rather insensitive to temperature, and for
elements with many energy levels with the same parity as the ground state, e.g., \ion{Fe}{2},
the increase in partition function further reduces the $T_{gas}$-sensitivity. 
It is only when $T_{gas} < T_{exc}$ that the fluxes become particularly sensitive to the gas temperature. 
(The upper energy levels of the CSE lines are given in Tables~\ref{tab:atomic} and \ref{tab:cse_lines}). 

A combination of the assumed temperature and density distributions is the most likely 
explanation for 
the discrepancy between observed and model mid-IR fluxes - 
noting that the \up[\ion{O}{1}\up] and \up[\ion{C}{2}\up] are in reasonable agreement. The
discrepancy appears to be a function of radius, being $3\times$ too high for the
chromosphere and wind base,  $2\times$ too high in the inner wind, and tending towards
agreement in the outer layers. In the inner wind region the density structure in the {\it Composite Model Atmosphere} 
has been interpolated, via a simple wind velocity model and the equation of continuity, and is not well constrained.
This could explain some of the flux discrepancies but not so readily the trend.

The inner $T_{gas}$-structure in the {\it Composite Model Atmosphere}
is derived from spatially resolved thermal radio emission. 
In the inhomogeneous atmosphere each line-of-sight through the stellar atmosphere
intersects material of 
different properties: some at the high temperatures responsible for the UV chromospheric emission,
and some much cooler and less ionized. The radio opacity is very sensitive to ionization 
($\kappa \propto n_e n_H$, $\kappa \propto n_e^2$) and hence
has a larger contribution from the hot material than does the forbidden \ion{Fe}{2} opacity
($\kappa \propto n_{Fe II}$). The over-estimation of fluxes from the excited $^4F$ term
suggests that the temperature of the bulk of the plasma where the chromosphere has its largest
filling factor is $< 2500$~K. Even though the radio brightness temperature inferred from the
VLA is significantly lower than previously expected (prior to 1998) from semi-empirical chromospheric
models, it appears 
that the radio brightness temperature is still greater than the temperature of the 
dominant gas component sampled by the \up[\ion{Fe}{2}\up]. 

As one moves away from the star, the filling factor of hot
chromospheric plasma decreases, and hence the difference between the mean temperature
inferred from the VLA radio interferometry and the bulk gas temperature decreases.
It is only by examining data from diagnostics with these different temperature and density
dependencies that we can hope to unravel the complex structures in
M supergiant atmospheres. We are now in an era where there are sufficient empirical constraints on the density, 
ionization, 
temperatures, and velocity fields that semi-theoretical models for the wind can be
investigated. From an observational standpoint the largest single improvement 
would be to have fully resolved, flux calibrated, line profiles for all the CSE emission lines 
obtained with good pointing accuracy. With such profiles, both the dynamic and thermodynamic constraints of 
these important cooling channels would be realized simultaneously. 

\section{Constraints on Wind Driving Mechanisms}

For these early M supergiants, radiation pressure on dust does not drive the
stellar outflows.  Most dust is located
far above the stellar surface \citep{bester_etal96} and the shells are not very opaque at the wavelengths of
the stellar flux peak.  It has not been shown that radiation pressure on atoms, ions, and molecules 
can drive the observed outflows. 
More likely candidates for driving the outflows include some form of pulsation \citep{lobel_dupree01} or 
MHD wave propagation, e.g., \cite{airapetian_etal00}.

The resolved TEXES profiles provide an estimate of the energy available to drive the stellar wind
which can be equated to that required to drive the observed mass outflow. 
The surface integrated energy flux required from the propagation of wave energy, neglecting wind radiative losses,
can be written as (see \citealt{holzer_macgregor85})
\begin{equation}
F_{wave}\left(R\right) = 4 \pi Z^2 R_\ast^2 \> V_{prop} \>{\cal C} \rho V_{turb}^2  \simeq
\dot{M} \left({GM_\ast\over{ZR_\ast}} + {V_\infty^2\over{2}}\right) =
{\dot{M} V^2_{esc}\over{2}} \left[{1\over{Z}} + \left({V_\infty\over{V_{esc}}}\right)^2\right]. 
\end{equation}
$V_{esc}$ is the surface escape speed $(\sim 65\>{\rm km\>s}^{-1}$) and 
${\cal C}$ is a factor of order unity that reflects line-of-sight projections and polarization 
of the wave motions \citep{jordan86}. 
Observationally it can be inferred that
the energy that drives mass-loss is mostly used to overcome the gravitational 
potential (i.e., $\propto V_{esc}^2$) with a small residual amount going into wind kinetic energy
(i.e., $\propto V_\infty^2$), so with $V_\infty \sim 10\>{\rm km\>s}^{-1}$, 
the ratio $(V_\infty/V_{esc})^2$ is small, i.e., $\simeq 0.024$.

Taking the atmospheric properties at the radius of the mid-point of the 
\up[\ion{Fe}{2}\up] contribution functions,  
we have estimates for $R^2 \rho(R)$, along with the measured value of $V_{turb}=12\>{\rm km\>s}^{-1}$
and $\dot{M}\simeq 4\times 10^{-6}\>{\rm M}_\odot\>{\rm yr}^{-1}$.
With these values the implied radial outward propagation velocity of the
wave energy is 
$V_{prop}\sim \> 5\>{\rm km\>s}^{-1}$ at $1.5R_\ast$. Blue-shifted emission
from a gas outflow with this velocity is not observed in either the UV or IR emission lines.

If the propagation speed corresponds to radial pulsations or acoustic waves then they 
are close to the sound speed, but they have not been observed.  Note that \cite{lobel_dupree01}
have infered non-radial pulsations with smaller amplitudes of order $1\>{\rm km\>s}^{-1}$.
If the energy propagation is Alfv\'enic then the implied magnetic field fluctuations 
have $\delta B = 0.4$\,Gauss. Since MHD fluctuations will damp when the amplitude 
approaches the radial field strength, $B \ge 0.4$\,G. The implied plasma $\beta = 8\pi P_{gas}/B^2$ is $\sim 1$, and the 
motions in the gas and magnetic field will be dynamically coupled. There are too many uncertainties
in our current knowledge of the radial dependence of atmospheric properties of
$\alpha$~Ori to be more definitive. The above arguments, namely the absence of emission
indicative of outward flows
at $1.5R_\ast$, suggests that either volume averaging of atmopsheric motions result in no
outflow signature, or that the wind energy flux is carried by
MHD fluctuations. The magnitude of the magnetic field and the order of the plasma $\beta$ 
suggest that wave damping remains a viable mechanism to drive mass-loss in Betelgeuse.

\section{CONCLUSIONS}

We present the first resolved spectroscopy of forbidden iron emission from M supergiants 
in the 20\,$\mu$m region.
The TEXES spectra allow us to examine the 
dynamics and thermodynamics of the extended atmospheres of early-type M supergiants. 
New accurate laboratory Ritz wavelengths from \cite{aldenius_johansson07}, 
and the accurate and reproducible absolute wavelength scales of the 
TEXES spectrograph allow the \up[\ion{Fe}{2}\up] 17.94\,$\mu$m and 
24.52\,$\mu$m emission lines to be scrutinized at the $1\>{\rm km\>s}^{-1}$ level,
which is also the accuracy at which stellar center-of-mass radial velocities of 
M supergiants are known.

Our results can be summarized as follows:
\begin{itemize}
\item The \up[\ion{Fe}{2}\up] emission lines are detected in all of our early M supergiant sample and
      the line-to-continuum flux ratios are consistent with the amount of circumstellar
      dust emission.  The lines widths show little variation within our sample.  
\item The $a\>^4F$ \up[\ion{Fe}{2}\up] emission profiles are spectrally resolved in the TEXES
      spectra, and we have now dynamically resolved the bulk cool plasma at the base 
      of the wind from the hot chromosphere. Although these lines are formed at the same 
      radial distances as the hot chromosphere observed in the ultraviolet, they have
      smaller intrinsic line widths, providing clues to the atmopsheric heating and mass-loss mechanisms.  
\item The emission cores of these \up[\ion{Fe}{2}\up] lines indicate that
      the lines are formed close to the star. The absence of blue-shifted 
      emission is in accord with low velocities expected in the line 
      forming region. 
\item The cool extended atmosphere has a radial velocity similar to
      that observed in hotter chromospheric UV (C~II\up])  diagnostics at 
      previous epochs. Neither component shows evidence of emission following
      the photospheric velocity fluctuations.
\item Detailed comparison of the observed fluxes of the \up[\ion{Fe}{2}\up] lines 
       from $\alpha$ Ori and a composite model 
      atmosphere are consistent with the
      view that Betelgeuse's extended atmosphere is {\em dominated} by cool gas.
      Early indications are that the bulk of the gas is even cooler than that
      inferred from the VLA radio interferometry, and that the
      filling factor of hot plasma declines throughout the first few stellar
      radii.       
\item We predict that spectrally resolved observations of the 
      25.99\,$\mu$m \up[\ion{Fe}{2}\up] line are likely to show a wind signature. This line is formed farther out than
      the $a\>^4F$ lines where the wind velocity is detectable with spectral resolutions of $R \ge 50,000$, 
      while not having too much
      contribution from the quasi-static region close to the star. This line was previously observed 
      at lower spectral resolution with ISO, but should be observable with EXES (similar to TEXES) on
      SOFIA \citep{richter_etal06}.
\item The \up[\ion{Fe}{2}\up] 17.94\,$\mu$m and 24.52\,$\mu$m line emission is co-spatial
      with the hot UV chromospheric and cool thermal radio continuum emission. The very 
      different sensitivities of these diagnostics to the thermal and ionization 
      structure are now beginning to constrain the filling factors of the different structural
      components.
\item The ground term \up[\ion{Fe}{1}\up] 24.04\,$\mu$m line in $\alpha$~Ori is narrower than the
      \up[\ion{Fe}{2}\up] which suggests that it is formed closer to the star where the
      turbulence is lower. The ratio of \up[\ion{Fe}{2}\up] to \up[\ion{Fe}{1}\up] fluxes
      indicates that iron is predominantly singly ionized in the extended atmosphere. 

\end{itemize}

In Appendix B we have  constructed an extended atmosphere and wind model for Betelgeuse 
but there are now sufficient empirical constraints to justify new theoretical thermodynamic 
and semi-empirical models that include 
a lower, more realistic, temperature boundary condition and are also constrained by the new 
mid- and far-IR and radio observations; however, this is beyond the scope of this 
present work.

\acknowledgments

This research was supported by NASA under ADP grant NNG04GD33G (GMH)
issued through the Office of Space Science, by an award issued by JPL/Caltech 
(No. 1275296, GMH) to support the interpretation of observations made with 
the Spitzer Space Telescope, which is operated by the Jet Propulsion Laboratory, 
California Institute of Technology, and NSF grant AST-0206367 (GMH, AB). 
MJR acknowledges grants NSF AST-0708074
and NASA NNG04GG92G. TEXES was built and the observations
funded by grants from NSF and the Texas Advanced Research Program.
NR is a Royal Swedish Academy of Sciences Research Fellow supported by a grant
from the Knut and Alice Wallenberg Foundation. 
Funding from Kungl. Fysiografiska S\"allskapet i Lund is acknowledged.
Based in part on observations obtained at the Gemini Observatory, which is operated by the
Association of Universities for Research in Astronomy, Inc., under a cooperative agreement
with the NSF on behalf of the Gemini partnership: the National Science Foundation (United
States), the Science and Technology Facilities Council (United Kingdom), the
National Research Council (Canada), CONICYT (Chile), the Australian Research Council
(Australia), MinistŽrio da Cincia e Tecnologia (Brazil) 
and Ministerio de Ciencia, Tecnolog'a e Innovaci—n Productiva  (Argentina).  This research was facilitated by NSF US-Sweden Cooperative Research Program 
grant INT-0318835 to the University
of Colorado and financial support from The Swedish
Foundation for International Cooperation in Research and Higher Education
(STINT), grant IG 2004-2074. 

This research has made use of the SIMBAD database, operated at CDS, 
Strasbourg, France, and used the DIRBE Point Source Photometry Tool, a 
service provided by the Legacy Archive for Microwave Background Data 
at NASA's Goddard Space Flight Center. We thank D. Gray for providing us
with photospheric radial velocity data, and for the referee, Dr. T. Ake,
for suggestions that improved the clarity of this paper.

\email{graham.harper@colorado.edu}.


%% copy editing.  Individual instruments can be provided in parentheses,
%% after the keyword, but they will not be verified.

{\it Facilities:} {\it Facilities:} \facility{IRTF (TEXES)}, \facility{Gemini:Gillett (TEXES)}, \facility{SOFIA (EXES)},
\facility{OVRO ()}, \facility{ISO ()}, \facility{KAO ()}, \facility{IRAS ()}



\clearpage

%% the file, uncomment the \plotone, \plottwo, and
%% If you need a layout that cannot be achieved with \plotone or
%% \plottwo, you can invoke the graphicx package directly with the
%% \includegraphics command or use \plotfiddle. For more information,
%% please see the tutorial on "Using Electronic Art with AASTeX" in the
%% documentation section at the AASTeX Web site,
%% http://www.journals.uchicago.edu/AAS/AASTeX.
%%
%% The examples below also include sample markup for submission of
%% supplemental electronic materials. As always, be sure to check
%% the instructions to authors for the journal you are submitting to
%% for specific submissions guidelines as they vary from

\begin{figure}
\epsscale{0.65}
\plotone{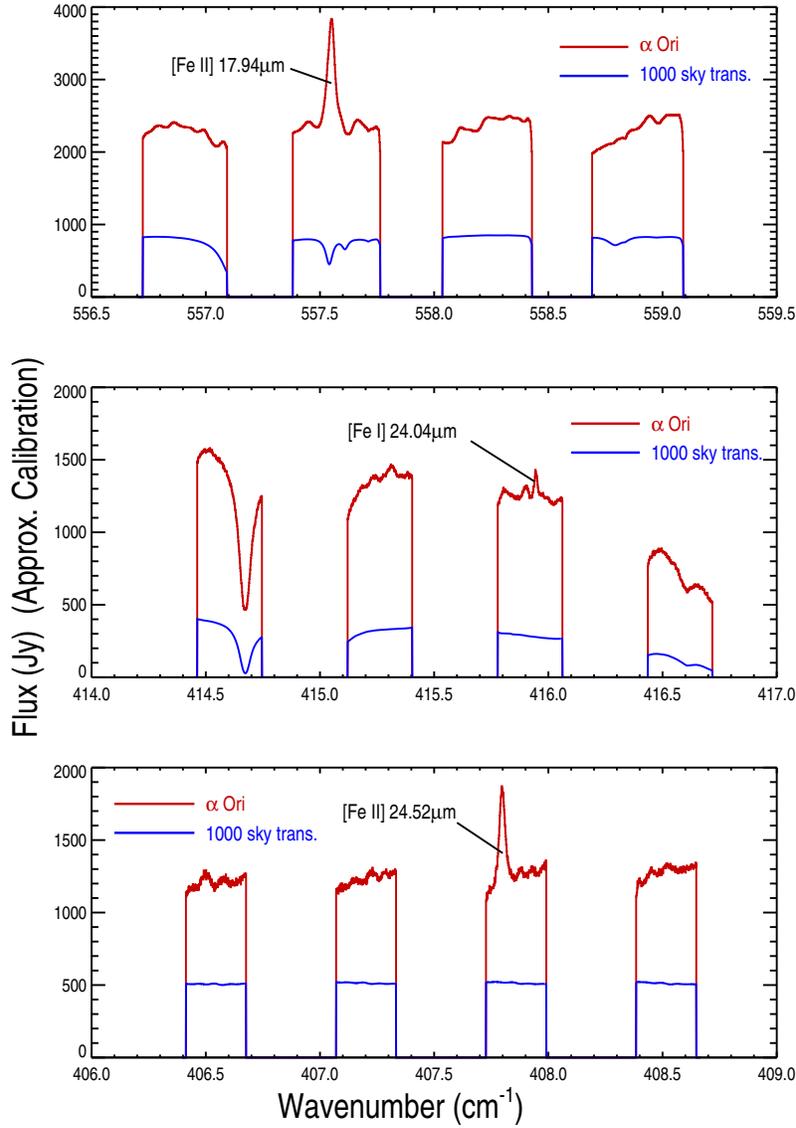}
\caption{TEXES observations of Betelgeuse. 
The stellar spectra with the flux calibration described in Appendix A
are shown in red. An estimate of the sky transmission is shown in blue.
The three emission lines are clearly visible and spectrally well resolved. 
The [Fe~II] lines are much stronger than [Fe~I] because 
Fe~II is the dominant ionization stage in the extended envelope. 
The telluric absorption features are used to establish a 
wavelength scale accurate to $<1 {\rm km\>s}^{-1}$. There is a telluric water line underlying the [Fe~II] 17.936\,$\mu$m line, 
while the shapes of the 
[Fe~II] 24.519 and [Fe~I] 24.042\,$\mu$m lines are 
not significantly affected by telluric water. The underlying stellar photosphere
also has shallow molecular features that give rise to the structured continuum.
\label{fig:texes_format} }
\end{figure}

\begin{figure}
\epsscale{1.0}
\plottwo{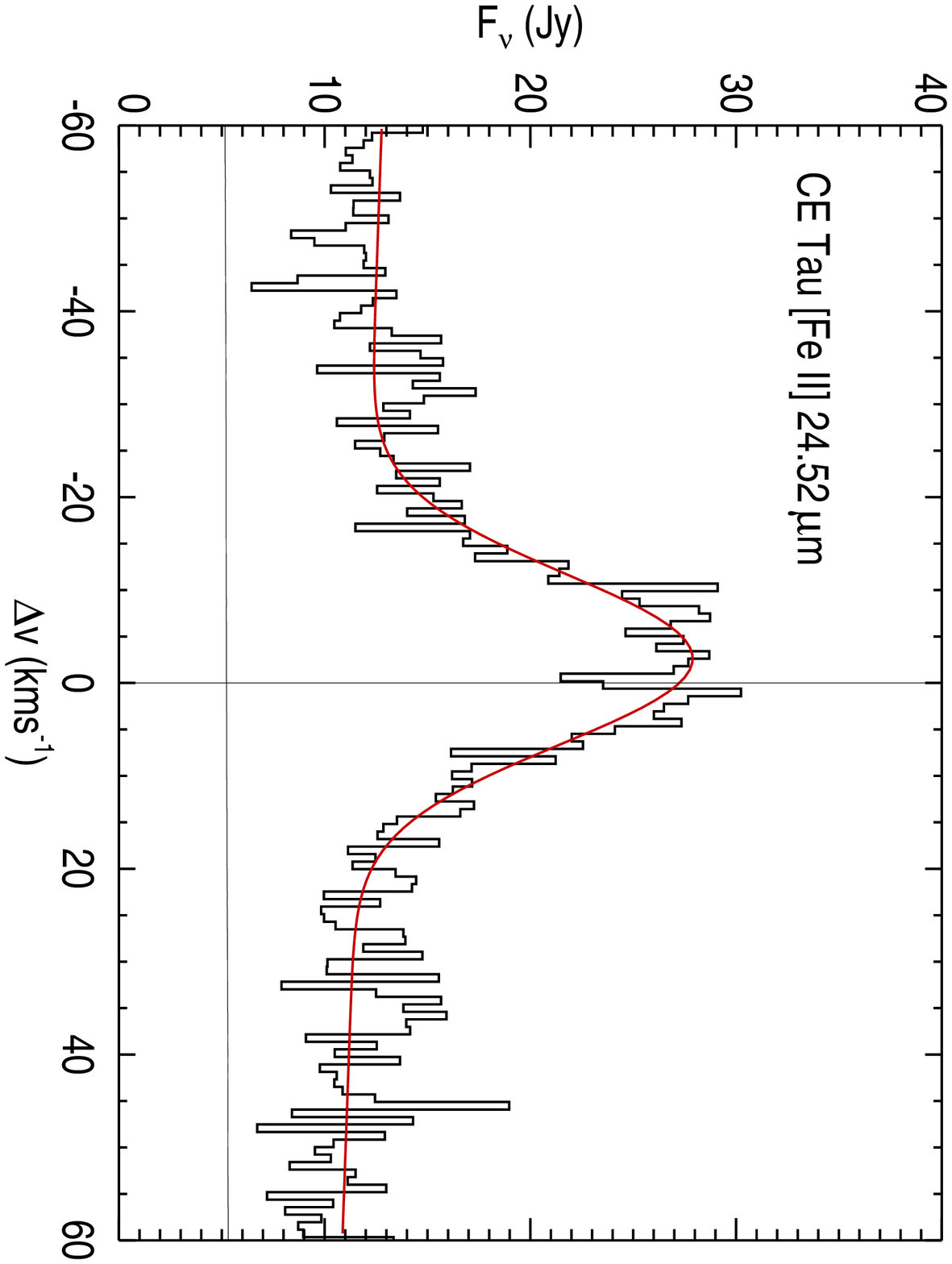}{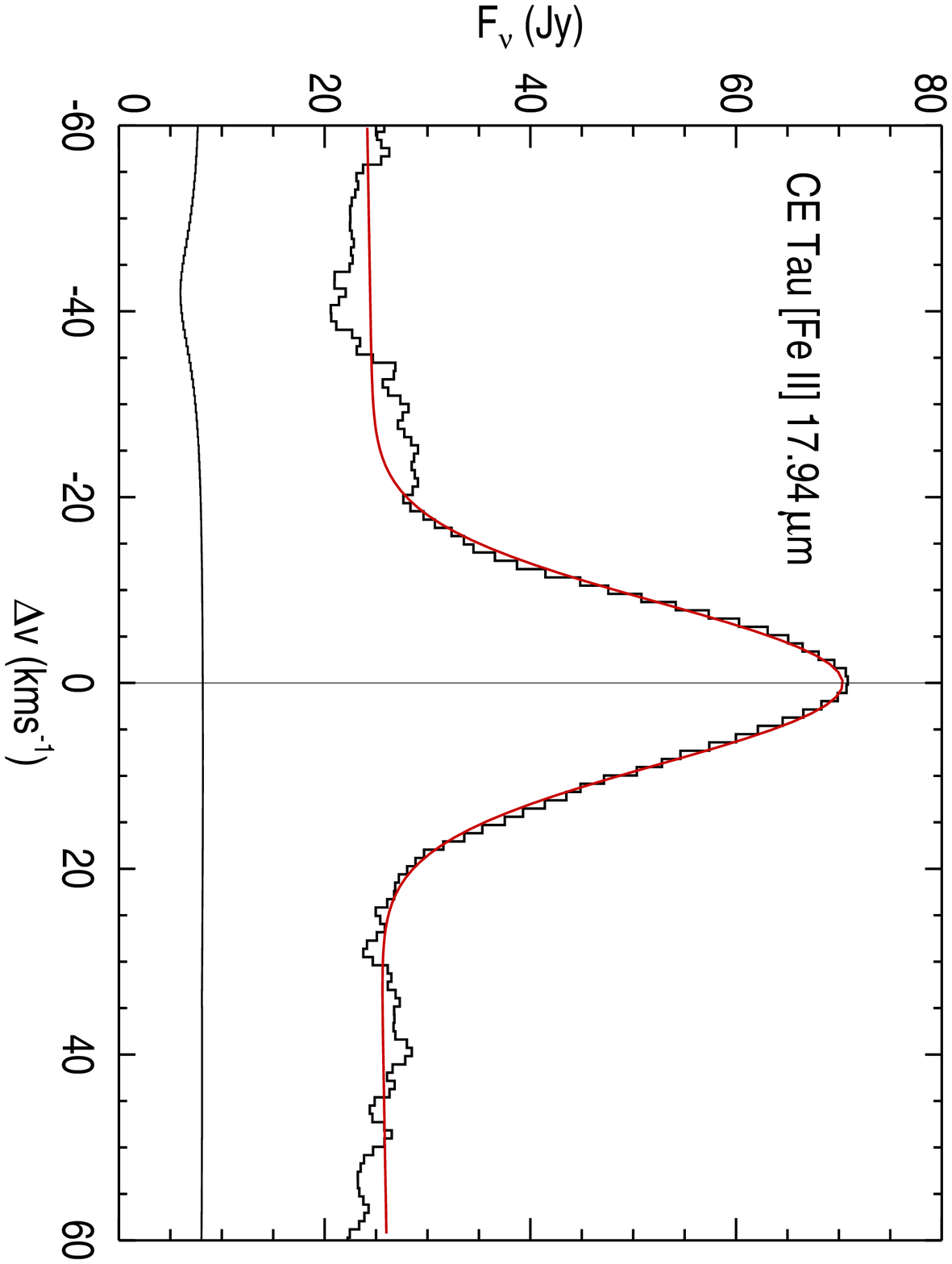}
\caption{\up[\ion{Fe}{2}\up] 17.94\,$\mu$m and 24.52\,$\mu$m emission lines in CE~Tau (M2~Iab). 
The abscissa is the Doppler velocity in the stellar rest frame, and the ordinate is the
flux which has been flat-fielded, but not put onto an absolute scale. The shape of the sky transmission
is shown beneath the spectrum.
CE~Tau is a spectral-type proxy for $\alpha$~Ori and the line centroid and widths for these
two stars are very similar. This star has weak (or absent) dust emission.\label{fig:cetau}}
\end{figure}

\newpage
\begin{figure}
\epsscale{1.0}
\plottwo{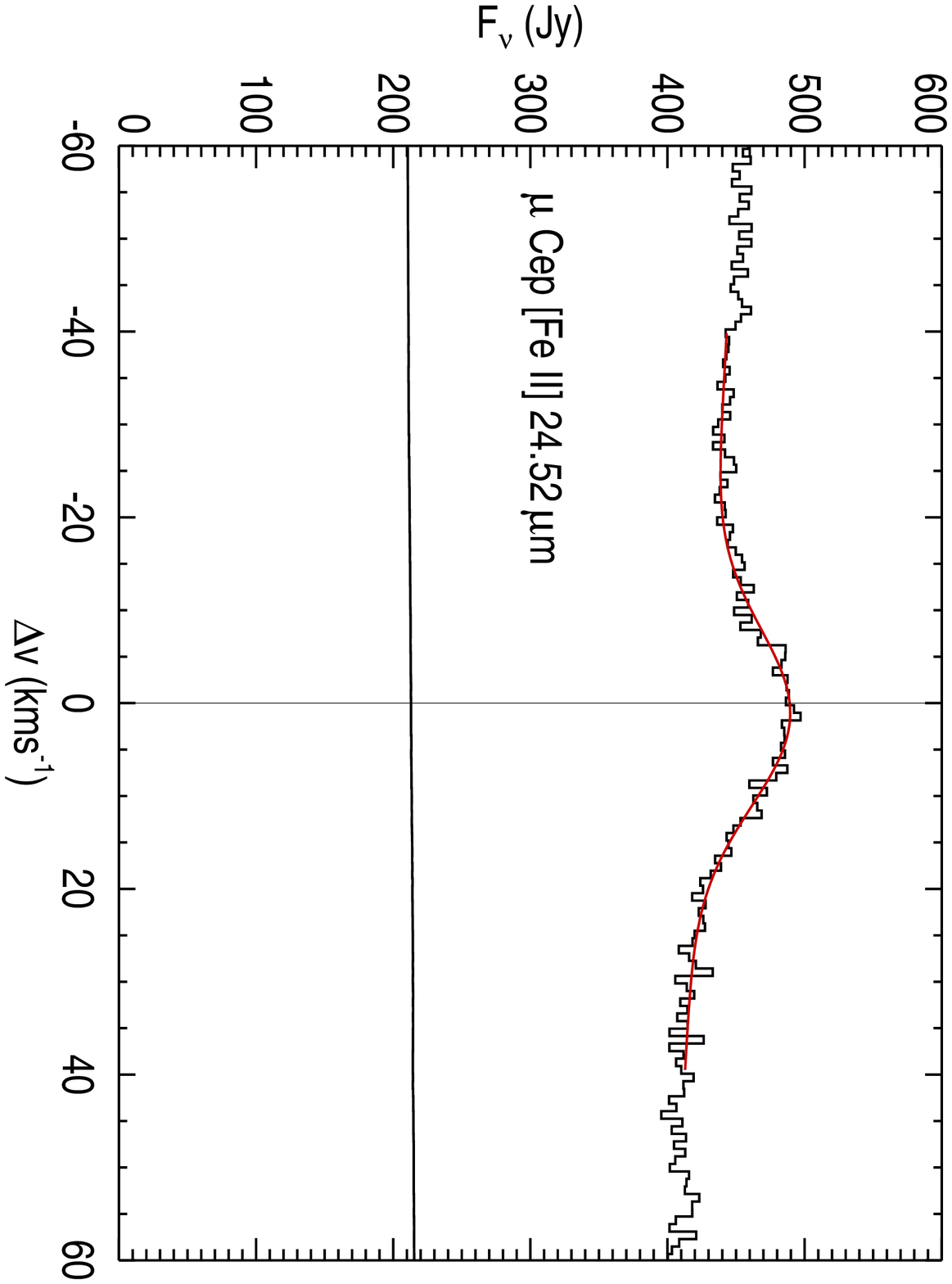}{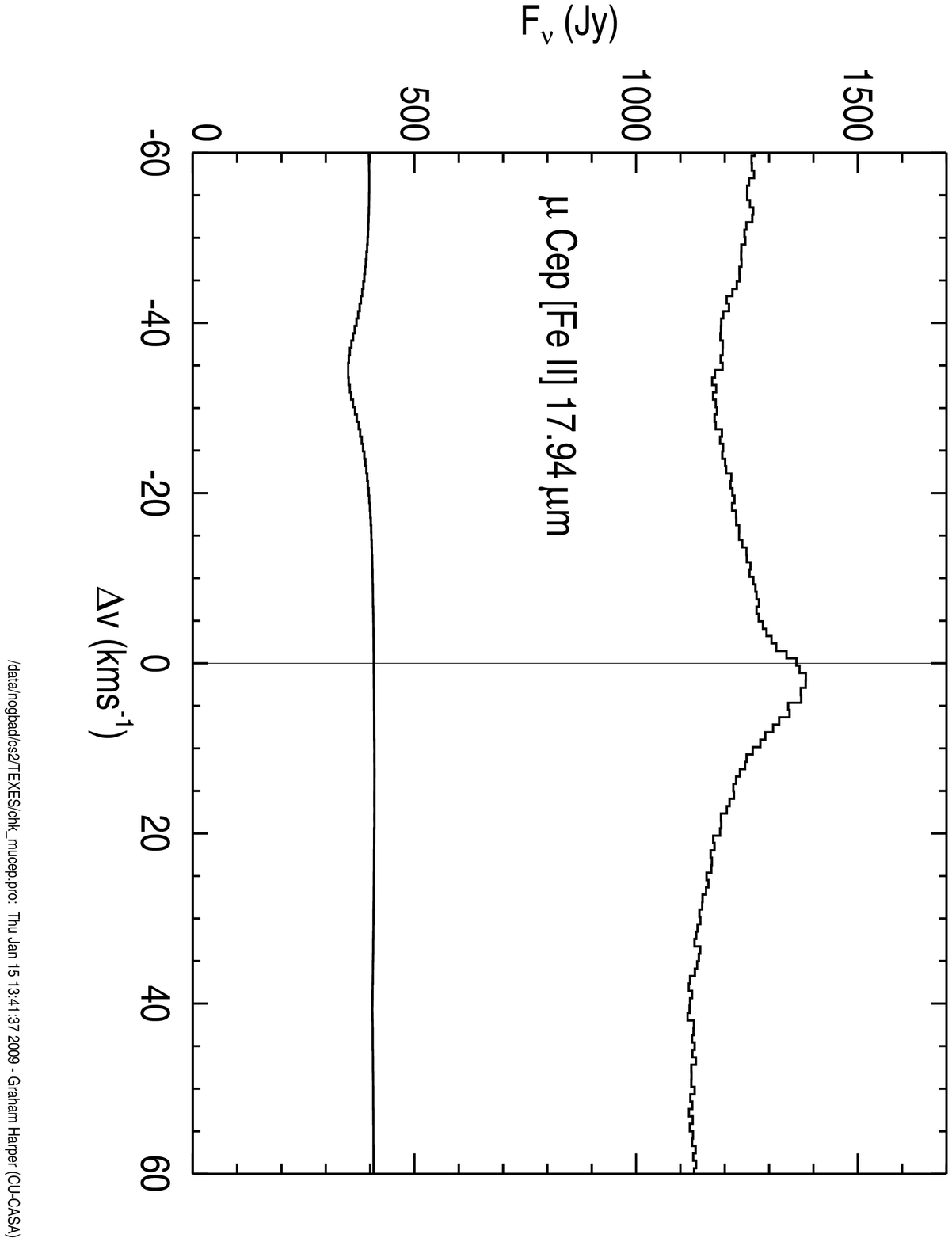}
\caption{\up[\ion{Fe}{2}\up] 17.94\,$\mu$m and 24.52\,$\mu$m emission lines in $\mu$~Cep (M2 Ia). 
The axes are the same as described in Fig.~\ref{fig:cetau}.
This supergiant has stronger silicate dust emission that $\alpha$~Ori which is a contributing 
factor to the low emission line to continuum ratio.
\label{fig:mucep} }
\end{figure}

\begin{figure}
\epsscale{0.8}
\plotone{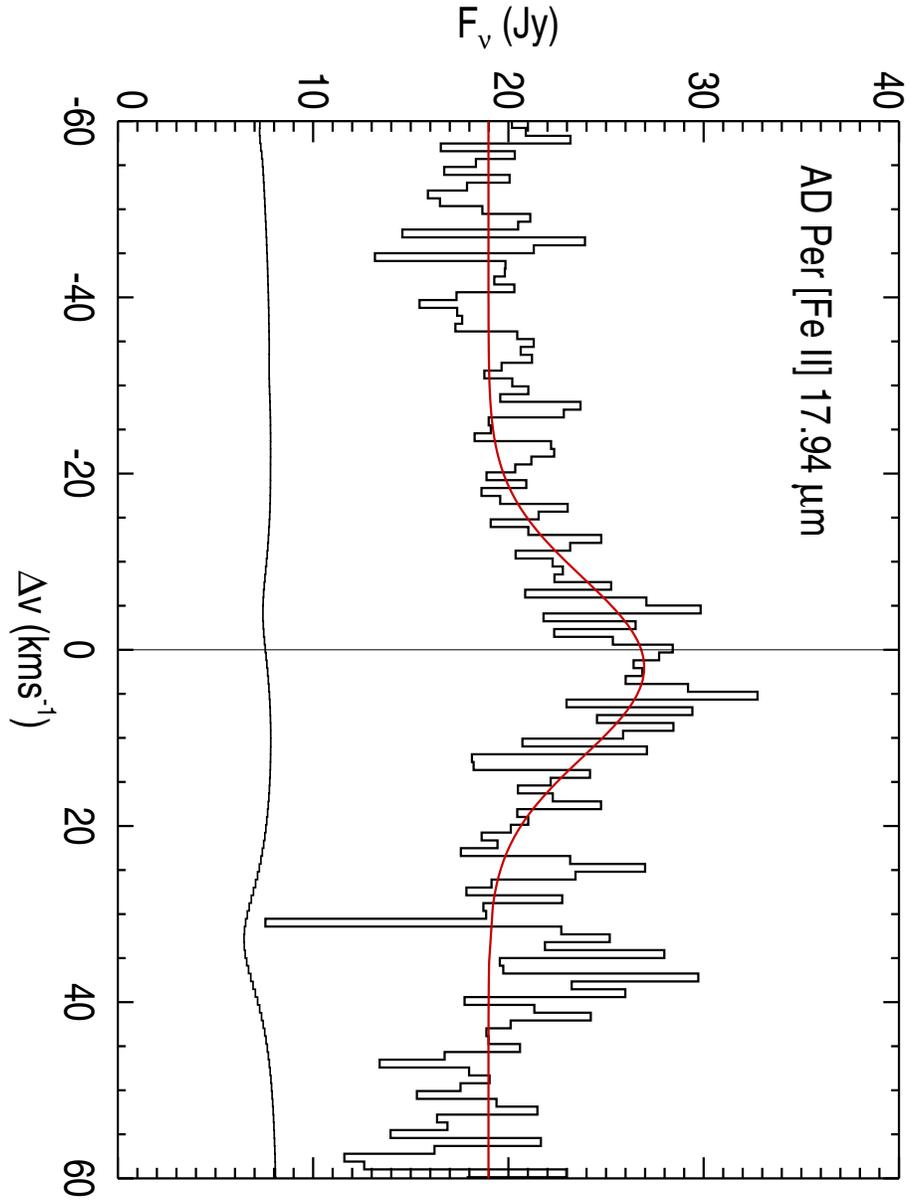}
\caption{\up[\ion{Fe}{2}\up] 17.94\,$\mu$m emission line in AD~Per. The axes are the same as described 
in Fig.~\ref{fig:cetau}. This is
the most distant M supergiant in the sample at $\sim 2$\,kpc and has a mass-loss rate
similar to nearby $\alpha$~Ori and $\alpha$~Sco which are $\sim 200$\,pc.\label{fig:adper}}
\end{figure}

\begin{figure}
\epsscale{0.7}
\plotone{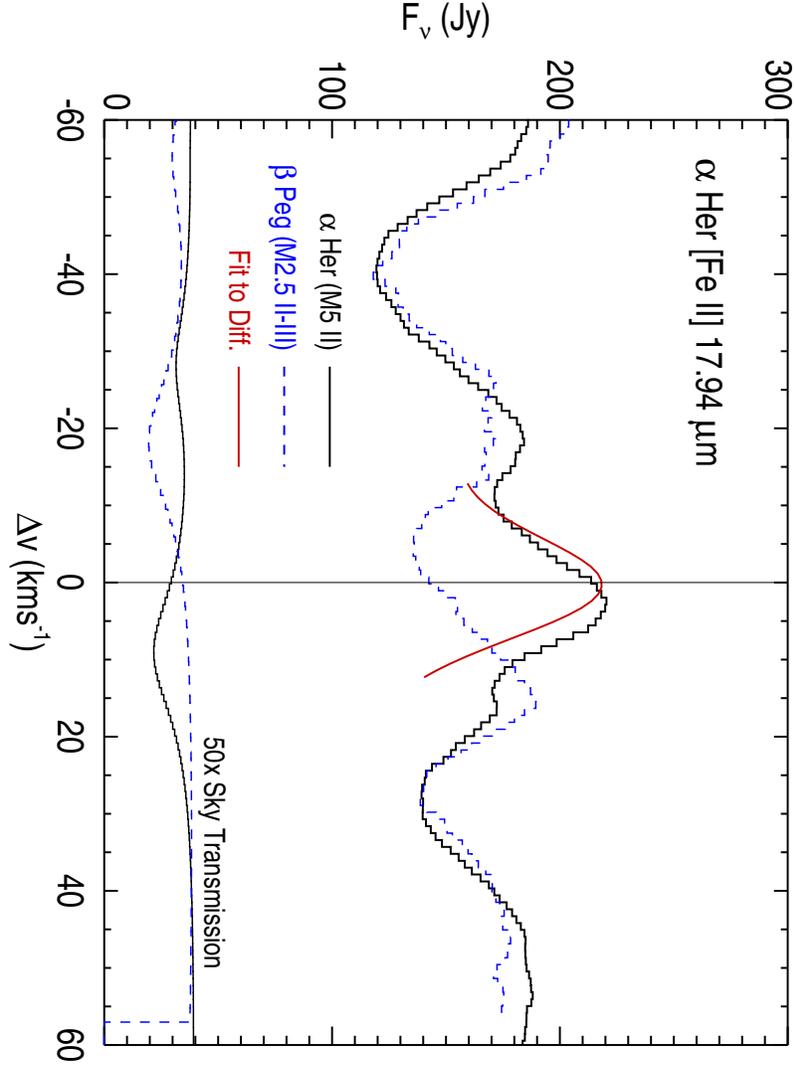}
\caption{Identification of \up[\ion{Fe}{2}\up] 17.94\,$\mu$m emission line in 
$\alpha^1$~Her (black) while $\beta$~Peg (blue) does not show an emission feature. 
Subtracting the scaled $\beta$~Peg spectrum from the $\alpha^1$~Her that has been flat-fielded
with NML~Cyg reveals emission at the
rest wavelength of $\alpha^1$~Her. A Gaussian fit to this difference spectrum is shown
in red and the properties are given in Table~\ref{tab:velocities}. The sky transmission for 
these two stars is also shown and significant additional uncertainties resulting
from the combined telluric correction are expected. Subsequent observations 
of the 24.52\,$\mu$m line have confirmed this detection of \up[\ion{Fe}{2}\up].
A color version of this figure is available in the electronic edition.\label{fig:aher}} 
\end{figure}

\begin{figure}
\epsscale{0.8}
\plotone{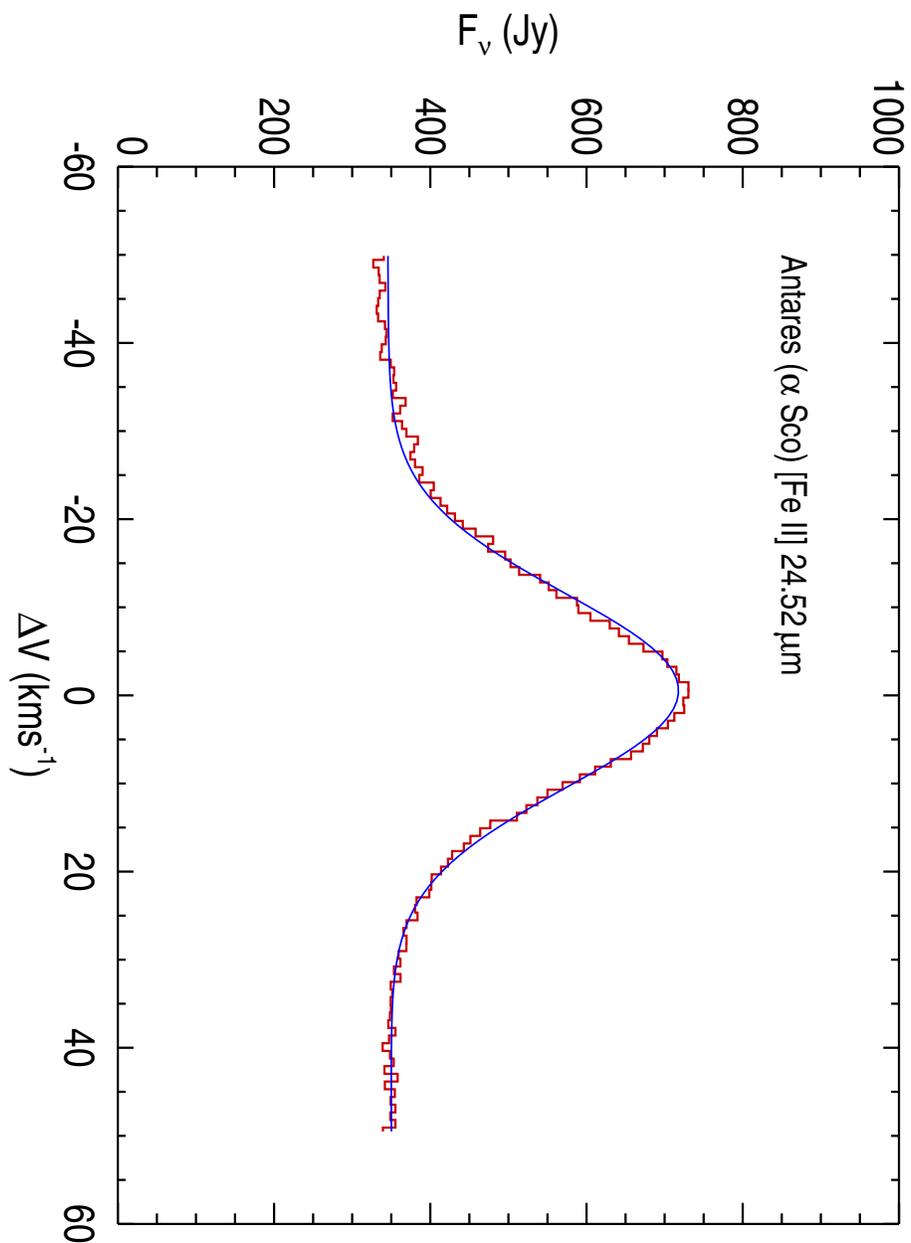}
\caption{\up[\ion{Fe}{2}\up] 24.52\,$\mu$m emission line in Antares. This spectrum
was obtained with Gemini-N during an engineering run on 2006 February 24. 
The axes are as described in Fig.~\ref{fig:cetau}. The
smooth profile is a Gaussian fit to the emission feature. No absolute wavelength scale 
was obtained for this spectrum.\label{fig:antares}}
\end{figure}

\begin{figure}
\epsscale{0.8}
\plotone{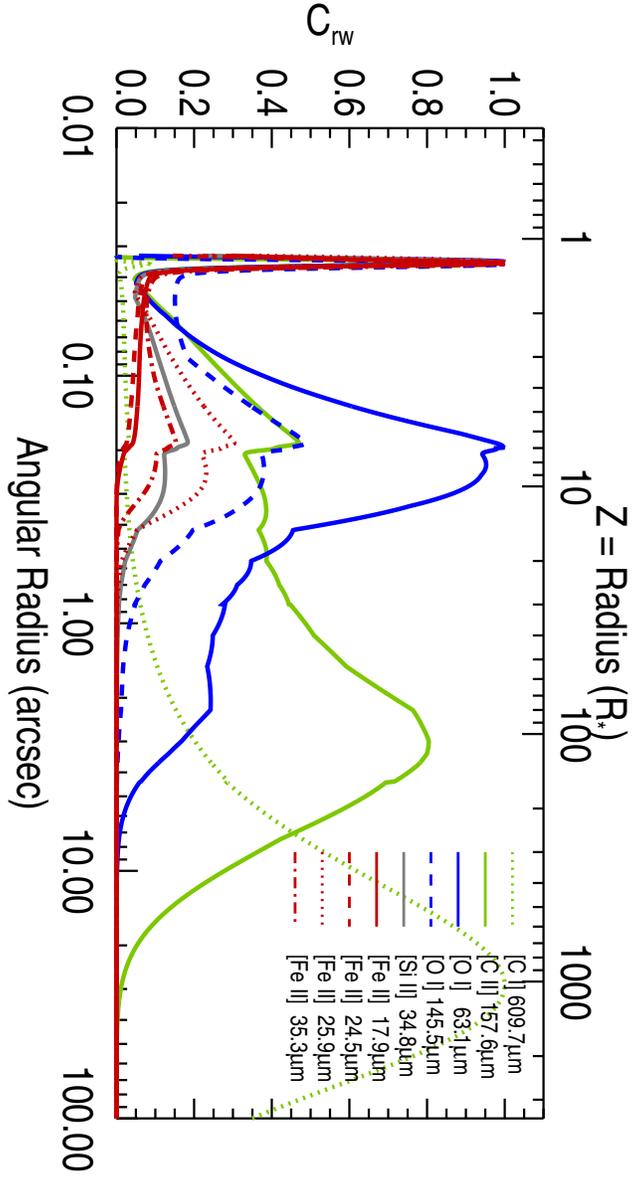}
\caption{Contribution functions for the emission lines observed in $\alpha$~Ori
with TEXES, {\it ISO} and the {\it KAO}. The area under each curve gives the
relative flux contribution. The narrow peak at $Z\sim 1.45$ is the
contribution from the relatively dense inhomogeneous region that 
includes the hot chromosphere.\label{fig:contrb}}
\end{figure}

%
%  TABLES
%

\clearpage
\begin{table}
\begin{center}
\caption{Radiative Atomic Data for Diagnostic Infrared [Fe~II] lines.}
\label{tab:atomic}

\begin{tabular}{lccrcrcc}
 \tableline\tableline
Species   & Wavelength\tablenotemark{a}  & Wavenumber & $E_{low}\>\>\>\>$  & $J_{low}$ & $E_{up}\>\>\>\>$     & $J_{up}$ & $A_{ji}$    \\ 
          & ($\mu$m)   & $({\rm cm}^{-1})$ & $({\rm cm}^{-1})$ &           & $({\rm cm}^{-1})$ &   & $({\rm
  s}^{-1})$
\tablenotemark{b}   \\  \tableline   
\underline{TEXES} & \multicolumn{7}{c}{}\\
\up[Fe II\up]     & 17.9360  & 557.5364 & 1872.6005 & 9/2 & 2430.1369 & 7/2 & $5.84\times 10^{-3}$   \\
\up[Fe II\up]     & 24.5192  & 407.8434 & 2430.1369 & 7/2 & 2837.9803 & 5/2 & $3.92\times 10^{-3}$   \\
\underline{ISO}   & \multicolumn{7}{c}{}\\
\up[Fe II\up]     & 25.9884  & 384.7868 & 0.0000   & 9/2 &  384.7868 & 7/2 & $2.13\times 10^{-3}$   \\
\up[Fe II\up]     & 35.3486  & 282.8963 & 384.7868 & 7/2 &  667.6830 & 5/2 & $1.57\times 10^{-3}$   \\\tableline
\end{tabular}
\tablenotetext{a}{Energy levels and wavelengths (vacuum) are from
\citet{aldenius_johansson07}.}
\tablenotetext{b}{Einstein A-values are from \cite{nussbaumer_storey88} and \cite{quinet_etal96}}
\end{center}
\end{table}

\clearpage
\begin{table}
\begin{center}
\caption{TEXES \up[Fe~II\up] and \up[Fe~I\up] Observation Summary for 2004 October 5,6,11, 2005 January 16,17, and
2005 December 9.}
\label{tab:obslog}
\vspace{0.3cm}
\begin{tabular}{lccccc}
 \tableline\tableline
Star         & Spectral-Type  & $V_{rad}$ & [FeII] 17.94\tablenotemark{a}    & [FeII] 24.52\tablenotemark{a}   & [FeI] 24.04\tablenotemark{a}    \\
             &              & (${\rm km\>s}^{-1}$) & ($\mu$m)  &  ($\mu$m)       &  ($\mu$m)         \\\tableline
$\mu$~Cep    & M2~Ia        & +19.4\tablenotemark{c}  & $\surd$     & $\surd$ & -        \\
$\alpha$~Sco & M1~Iab       & -3.5\tablenotemark{c}    & -           & $\surd$ & -        \\
$\alpha$~Ori & M2~Iab       & +20.7\tablenotemark{d}     & $\surd$     & $\surd$ & $\surd$  \\
CE~Tau      & M2~Iab        & +22.8\tablenotemark{c}     & $\surd$     & $\surd$ & -       \\
AD~Per       & M2.5~Iab     & -44 \tablenotemark{e}      & $\surd$     &  -      & -       \\
$\alpha$~Her & M5~II        & -33.1 \tablenotemark{c}    & $\surd$\tablenotemark{b} &  -      & X       \\
$\beta$~Peg  & M2.5~II-III  & +9.1\tablenotemark{c}      & X           &  -      & -       \\
Mira         & M7~III       & +63.5\tablenotemark{c}     & X           &  X      & -       \\
$\zeta$~Aur  & K4~Ib-II     & Binary\tablenotemark{c}    & X           &  -      & -       \\
$\alpha$~Tau & K5~III       & +54.3\tablenotemark{c}     & X           & X       & -       \\ \tableline
\end{tabular}
\tablenotetext{a}{Key: detection ($\surd$), non-detection (X), and not observed (-).}
\tablenotetext{b}{As judged by comparison with $\beta$~Peg, see Fig \ref{fig:aher}. 
\up[\ion{Fe}{2}\up] emission was subsequently confirmed at 24.52$\mu$m.} 
\tablenotetext{c}{\cite{barbier-brossat_fignon00}.}
\tablenotetext{d}{Mean of \cite{jones28} and \cite{sanford33}.}
\tablenotetext{e}{Mean of \cite{barbier-brossat_fignon00} and \cite{mermilliod_etal08}.}
\end{center}
\end{table}

\clearpage
\begin{table}
\begin{center}
\caption{Line Centroid Velocities and Doppler Widths ($\pm 1\sigma$\tablenotemark{a}) of the \up[\ion{Fe}{2}\up] 
Emission Lines. }

\label{tab:velocities}
\vspace{0.3cm}
\begin{tabular}{lcccc}
 \tableline\tableline
Star  & Spectral-Type  & $V_{rad}$ & $V_{cent}$ \tablenotemark{b} & $V_{Dopp}$(Obs)\tablenotemark{c}\\
  &   & (${\rm km\>s}^{-1}$) & (${\rm km\>s}^{-1}$) & (${\rm km\>s}^{-1}$) \\\tableline
$\mu$~Cep    & M2~Ia          & +19.4     & $1.7\pm 0.2$ & $13.4 \pm 0.4$  \\   
$\alpha$~Sco & M1~Iab         & -3.5      & No WaveCal   & $15.6 \pm 0.3$  \\
$\alpha$~Ori & M2~Iab         & +20.7     & $0.0\pm 0.6$ & $12.5 \pm 0.8$  \\
CE~Tau       & M2~Iab         & +22.8     & $0.0\pm 0.1$ & $12.0 \pm 0.2$  \\
AD~Per       & M2.5~Iab       & -44       & $2.0\pm 1.5$ & $14.4 \pm 1.4$  \\
$\alpha$~Her & M5~II          & -33.1     & $1.7\pm 0.5$ & $ 9.2 \pm 0.8$\tablenotemark{d} \\\tableline
\end{tabular}
\tablenotetext{a}{$1\sigma$ are either the formal uncertainty of the Gaussian profile fit, or the
dispersion of from multiple epoch measurements.}
\tablenotetext{b}{Centroid velocities ($V_{cent}$) are with respect to the
adopted stellar center-of-mass radial velocities, $V_{rad}$.}
\tablenotetext{c}{Observed Doppler widths, $V_{Dopp}$(Obs), are defined in terms of the Full Width at Half Maximum: 
FWHM$=1.665 V_{Dopp}$(Obs), and are uncorrected for insturmental line broadening.}
\tablenotetext{d}{This is a heavily blended feature, see Fig \ref{fig:aher}, and the
uncertainties are dominated by systematic errors for this star.} 
\end{center}
\end{table}

\clearpage
\begin{table}
\begin{center}
\caption{Properties of $\alpha$~Ori's TEXES \up[\ion{Fe}{2}\up] emission lines.} 
\label{tab:alphaori}
\vspace{0.3cm}
\begin{tabular}{lcccc}
 \tableline\tableline
Date         & Flux\tablenotemark{a}                 & $V_{cent}$            & $V_{Dopp}$(Obs) \\
UT           & ($10^{-19}\>{\rm W\>cm}^{-2}$) & (${\rm km\>s}^{-1}$)  & (${\rm km\>s}^{-1}$) \\\tableline
\underline{\ion{Fe}{2} 24.52\,$\mu$m} &          &                       &                 \\
2004 Oct 05  & $6.2\pm 0.1$ & No WaveCal    & $13.2\pm 0.2$ \\
2004 Oct 06  & $6.2\pm 0.2$ & No WaveCal    & $12.4\pm 0.5$ \\
2004 Oct 11  & $5.7\pm 0.1$ & $+0.7\pm 0.1$ & $12.8\pm 0.3$ \\
2005 Jan 16  & $6.0\pm 0.1$ & $-1.0\pm 0.1$ & $13.3\pm 0.3$ \\
2005 Dec 09  & $6.0\pm 0.1$ & $-0.4\pm 0.1$ & $12.9\pm 0.2$ \\
\underline{\ion{Fe}{2} 17.94\,$\mu$m} &          &                       &                 \\
2004 Oct 05  & $16.5\pm 0.5$ &  $0.8\pm 0.2$ & $12.0\pm 0.4$ \\
2005 Jan 16  & $16.1\pm 0.1$ &  $0.3\pm 0.1$ & $11.7\pm 0.1$ \\
2005 Dec 07  & $16.0\pm 0.2$ &  $0.6\pm 0.1$ & $11.4\pm 0.2$ \\
\underline{\ion{Fe}{1} 24.04\,$\mu$m} &          &                       &                 \\
2004 Oct 06  & $ 0.77\pm 0.03$ &  $-2.7\pm 0.2$ & $6.5\pm 0.3$ \\ \tableline
\end{tabular}
\tablenotetext{a}{Flux is the emission measured above the local
contiunuum and all $1\sigma$ uncertainties are from the formal fits to a Gaussian
profile.}
\end{center}
\end{table}

\clearpage
\begin{table}
\begin{center}
\caption{Properties of Betelgeuse's Infrared [\ion{Fe}{1}\up] and \up[\ion{Fe}{2}\up] lines.}
\label{tab:feii_observed}
\begin{tabular}{lccccc}
 \tableline\tableline
Ion & Wavelength     & $V_{cent}$     & $V_{turb}$           & Flux & Flux (model)         \\ 
    & (Vac. $\mu$m)  & $({\rm km\>s}^{-1})$ & $({\rm km\>s}^{-1})$   & $({\rm W\>cm}^{-2})$ & $({\rm W\>cm}^{-2})$ \\  \tableline   
\underline{TEXES}\tablenotemark{a} & \multicolumn{5}{c}{}\\
\up[Fe II\up] & 17.9360 & $+0.5\pm 0.2$  & $12\pm 0.1$   & $1.6 \pm 0.1 \times 10^{-18}$ & $5.4\times 10^{-18}$  \\
\up[Fe II\up] & 24.5192 & $+0.0\pm 0.4$  & $13\pm 0.2$   & $5.9 \pm 0.2 \times 10^{-19}$ & $1.8\times 10^{-18}$  \\
\up[Fe I\up]  & 24.0423 & $-2.7\pm 0.2$  & $6.5\pm 0.4$  & $7.7 \pm 0.3 \times 10^{-20}$ & \tablenotemark{c}$7.4\times 10^{-20}$ \\
\underline{ISO}\tablenotemark{b} & \multicolumn{5}{c}{}\\
\up[Fe II\up] & 25.9884 & \multicolumn{2}{c}{unresolved} & $2.8 \pm 0.1 \times 10^{-18}$ & $4.4\times 10^{-18}$ \\
\up[Fe II\up] & 35.3486 & \multicolumn{2}{c}{unresolved} & $8.3 \pm 0.3 \times 10^{-19}$ & $1.4\times 10^{-18}$ \\
\up[Fe I\up]  & 34.7133 & \multicolumn{2}{c}{unresolved} & $< 5 \times 10^{-20}$         & \tablenotemark{c}$2.2\times 10^{-20}$ \\ \tableline
\end{tabular}
\tablenotetext{a}{TEXES fluxes are from this work.}
\tablenotetext{b}{{\it ISO} fluxes use the normalization described in Appendix A.}
\tablenotetext{c}{Assuming $A_{Fe I} = 10^{-2}A_{Fe}$.}
\end{center}
\end{table}

\clearpage
\begin{deluxetable}{lccc}
\rotate
\tablecaption{OVRO 100~GHz Radio fluxes for $\alpha$~Sco, $\alpha$~Ori, and $\alpha$~Her.
\label{tab:ovro}}
\tablehead{ \colhead{} & \colhead{$\alpha$~Sco} &  \colhead{$\alpha$~Ori} & \colhead{$\alpha$~Her} }
\startdata
Date           & 2004 Mar   30 &  2003 Nov 9      & 2004 Mar 30   \\
Exposure Time (Hour) &   2     &    2             &   5   \\
$\alpha$ (J2000)     & 16h 29\arcmin\, 24.492\arcsec & 5h 55\arcmin\, 10.322\arcsec & 17h 14\arcmin\, 38.862\arcsec \\
$\sigma$($\alpha$)   &  $\pm$ 0.002\arcsec & $\pm$ 0.005\arcsec &  $\pm$ 0.002\arcsec \\
$\delta$ (J2000)    &  $-26\degr$ 25\arcmin\, 54.676\arcsec & $7\degr$ 24\arcmin\, 25.302\arcsec & $14\degr$ 23\arcmin\, 25.611\arcsec \\
$\sigma$($\delta$)  & $\pm$ 0.005\arcsec & $\pm$ 0.034\arcsec & $\pm$ 0.002\arcsec \\
Antenna Configuration  & E     &    L    &    E \\
Beam Size, Position Angle & $8.6 \times 4.3$\arcsec, $-4.0\degr$ & $15.4\times 5.0$\arcsec, $-11.0\degr$ & $4.9\times 4.1$\arcsec,  $+75.8\degr$\\
100 GHz Flux \& 1$\sigma$ (mJy) &  $ 70 \pm 1.8$  & $80.4\pm 3.7$  & $19\pm 0.49$ \\
250~GHz Flux\tablenotemark{a} \& 1$\sigma$ (mJy) & $345\pm 34$ & $351\pm 25$ & $104\pm 10$ \\
\enddata
\tablenotetext{a}{250~GHz fluxes are from \cite{altenhoff_etal94}.}
\end{deluxetable}

\clearpage
\begin{table}
\begin{center}
\caption{Atomic Data for CSE emission lines, and adopted abundances for $\alpha$~Ori.}
\label{tab:cse_lines}
\begin{tabular}{lccccl}
 \tableline\tableline
Transition     & Wavelength &  $A_{ji}$\tablenotemark{a}        & Abundance & $E_{up}$  & Source for Abundance\\ 
               & ($\mu$m)   & (${\rm s}^{-1}$) & (Rel. to H) & $({\rm cm}^{-1})$ &  \\  \tableline   
\up[Fe~I\up]   & 24.04      & 2.51d-03         & 3.0d-05   & 415.933  & \cite{carr_etal00}   \\
\up[Fe II\up]  & 25.99      & 2.13d-03         & 3.0d-05   & 384.790  & \cite{carr_etal00}   \\
\up[Fe II\up]  & 35.35      & 1.57d-03         & 3.0d-05   & 667.683  & \cite{carr_etal00}   \\
\up[O I\up]    & 63.18      & 8.91d-05         & 6.3d-04   & 158.265  & \cite{lambert_etal84}  \\
\up[O I\up]    & 145.5      & 1.75d-05         & 6.3d-04   & 226.977  & \cite{lambert_etal84}  \\
\up[Si~II\up]  & 34.81      & 2.13d-04         & 3.8d-05   & 287.24   & \cite{rodgers_glassgold91}   \\
\up[C~I\up]    & 609.7      & 7.88d-08         & 2.5d-04   & 16.40    & \cite{lambert_etal84}   \\
\up[C~II\up]   & 157.7      & 2.30d-06         & 2.5d-04   & 63.42    & \cite{lambert_etal84}   \\  \tableline     
\end{tabular}
\tablenotetext{a}{Einstein A-values are from NIST\tablenotemark{\dag} except for Fe~I \citep{brown_evenson95}.}
\tablenotetext{\dag}{Ralchenko, Yu., Kramida, A.E., Reader, J. and NIST ASD Team (2008). NIST Atomic Spectra Database (version 3.1.4), 
[Online]. Available: http://physics.nist.gov/asd3 [2008, April 16]. National Institute of Standards and Technology, Gaithersburg, MD.}
\end{center}
\end{table}

\clearpage
\begin{table}
\begin{center}
\caption{Other observed forbidden line fluxes\tablenotemark{a} for $\alpha$~Ori and computed flxues
from the {\it Composite Model Atmosphere} described in Appendix B.}
\label{tab:fluxes_other}
\begin{tabular}{lccl}
 \tableline\tableline
Transition        & Flux (observed)      & Flux (model)         & Reference for Observed Fluxes      \\ 
                  & $({\rm W\>cm}^{-2})$ & $({\rm W\>cm}^{-2})$ &                      \\  \tableline   
                  &                      &                      &                      \\
\up[O I\up] 63.18\,$\mu$m   & $2.4 \pm 0.2 \times 10^{-18}$    & $8.9\times 10^{-19}$ & \cite{haas_glassgold93}  \\
                             & $1.1 \pm 0.2 \times 10^{-18}$    &                      & \cite{haas_etal95}\tablenotemark{b}  \\
                             & $1.9 \pm 0.1 \times 10^{-18}$    &                      & {\it ISO} \cite{castro_carrizo_etal01}  \\
\up[O I\up] 145.5\,$\mu$m   & $11  \pm 4   \times 10^{-20}$    & $4.8\times 10^{-20}$ & \cite{haas_etal95}  \\
                             & $2.7 \pm 0.5 \times 10^{-20}$    &                      & \cite{castro_carrizo_etal01}    \\
                             & $5           \times 10^{-20}$    &                      & \cite{barlow99} (\& Priv. Comm)  \\
\up[Si~II\up] 34.81\,$\mu$m & $0.93 \pm 0.08 \times 10^{-18}$  & $1.5\times 10^{-18}$ & {\it ISO} This paper  \\
                             & $0.94 \pm 0.37 \times 10^{-18}$  &                      & \cite{haas_glassgold93}\\
\up[C~II\up] 157.7\,$\mu$m  & $1.1 \pm 0.1  \times 10^{-19}$   & $0.7\times 10^{-19}$ & \cite{barlow99}   \\     
                             &$1.2 \pm 0.1  \times 10^{-19}$   &                      &\cite{castro_carrizo_etal01}\tablenotemark{b}\\
                             &                                  &                      &           \\ \tableline
\end{tabular}
\tablenotetext{a}{Assuming each ion is the dominant ionization state.}
\tablenotetext{b}{Off-source emission reported at 50\% level.}
\end{center}
\end{table}

\clearpage
\appendix
\section{Approximate Absolute Flux Calibration for the TEXES $\alpha$~Ori Spectra}

The analysis of M supergiant atmospheric dynamics can be made without
an absolute flux calibration of the TEXES spectra, but to explore the full thermodynamic 
diagnostic potential of the \up[\ion{Fe}{2}\up] 24.52\,$\mu$m and 17.94\,$\mu$m, and
\up[\ion{Fe}{1}\up] 24.04\,$\mu$m emission lines requires absolute flux calibrated spectra. 
This enables a comparison of flux predictions from model atmospheres with
observations from airborne and space observatories, i.e., \up[\ion{Si}{2}\up] and \up[\ion{O}{1}\up] emission 
detected with {\it KAO}, and \up[\ion{C}{2}\up] and \up[\ion{Fe}{2}\up] emission in {\it ISO} spectra.
Here we describe an approximate flux calibration of the TEXES spectra for $\alpha$~Ori
which has several independent mid-IR flux measurements that can be used to calibrate and 
correct for the TEXES slit losses.

In this Appendix we bring the TEXES spectra to an absolute scale by adopting the shape of 
published {\it ISO} Short Wavelength Spectrometer (SWS) spectra over the 16.5-26.5\,$\mu$m wavelength region 
and scale the continuum flux at 25\,$\mu$m to a value derived from a combination of fluxes
from color-corrected photometry from the Diffuse Infrared Background Experiment (DIRBE) on the
{\it Cosmic Background Explorer (COBE)} satellite \citep{boggess_etal92}, 
color-corrected {\it InfraRed Astronomical Satellite (IRAS)} photometry, and
the cryogenic grating spectrometer (CGS) on the {\it KAO} 
\citep{haas_glassgold93}. These are also checked against 8-13\,$\mu$m UKIRT CGS3 spectrophotometry
\citep{monnier_etal98}.

$\alpha$~Ori's mid-IR spectra contains emission from close to the
star and from spatially extended optically thin silicate dust emission. The TEXES, {\it ISO}, 
{\it IRAS}, {\it KAO}-CGS, and DIRBE 
observations all have different entrance apertures and beam sizes and as 
a consequence the TEXES spectra are scaled with differential 
corrections that account for the different TEXES slit losses for 
the extended dust and point source stellar emission. The individual steps are outlined below.

\subsection{{\it ISO} Spectrometers}

To derive color-corrected fluxes for the DIRBE and {\it IRAS} photometry the
spectral shape and system responses are required across each photometric passband. 
To find the color-corrections  $K_\lambda$, where the observed color-corrected flux is given by
$F_{obs}=F_{DIRBE,IRAS}/K_\lambda$, we adopted {\it ISO} spectra and MARCS models 
(see Table \ref{tab:dirbe}). 

Three different reductions of $\alpha$~Ori's {\it ISO} SWS spectra, obtained with a grating resolution of $R\sim 1000$  at scan speed \#4, 
from \cite{justtanont_etal99}, \cite{verhoelst_etal06}, and \cite{sloan_etal03} were used
to derive $K_\lambda$ for the DIRBE 3.5\,$\mu$m, 4.9\,$\mu$m, 12\,$\mu$m, and 25\,$\mu$m and {\it IRAS} 12, and 25\,$\mu$m fluxes. 
The LWS grating spectrum\footnote{Kindly provided by M. Barlow.} which was
obtained two days before the end of the {\it ISO} mission \citep{barlow99} was
used to find $K_\lambda$ for the 60 and 100\,$\mu$m fluxes.

The \up[\ion{Fe}{2}\up] TEXES observations correspond to the {\it ISO}
SWS band 3 which has a $14\times 27\arcsec$ aperture. Betelgeuse is a bright IR source 
and provides an {\it ISO} calibration challenge in addition 
to {\it ISO}'s known systematic flux calibration errors \citep{verhoelst_etal06}.  
This is a reason why the published flux spectrum values differ by as much as 40\%. In the following we
boot-strap the {\it ISO} flux calibration using the color-corrected photometry.

\subsection{DIRBE Photometry}

We processed the DIRBE photometry from the Calibrated Individual
Observations in a fashion similar to \cite{smith03} and
\cite{smith_etal04}.  Flux outliers caused by cosmic rays coincident with, and off, the  
source position were rejected. We did not use IR databases to reject 
certain scan directions to avoid potential source confusion which can be important at shorter 
wavelengths. The color-uncorrected fluxes are given in Table~\ref{tab:dirbe}.
These are essentially identical to those found by \cite{smith03} and \cite{smith_etal04}. 
There are typically about 300 measurements in each band.

At 25\,$\mu$m and 60\,$\mu$m the mean of the individual flux error estimates 
is similar to the $1\sigma$ standard deviation of the total dataset, 
while at 12\,$\mu$m the $1\sigma$ standard deviation is 
three times the mean error hinting at intrinsic short term
variability during this epoch.
Over the 10 months of cryogenic DIRBE observations there was a
7\% increase in flux at 12\,$\mu$m and a 3\% increase at 25\,$\mu$m.
Note that the standard deviation of the mean is an order of magnitude 
smaller than the standard deviation of the individual measurements.

To convert the DIRBE photometry into color-corrected 
fluxes the spectral distributions across each band were combined with the 
spectral response curves from \cite{hauser_etal98}. The color-corrected fluxes
are given in Table~\ref{tab:dirbe} 

\subsection{{\it IRAS} Photometry and Low Resolution Spectrometer}

Betelgeuse had {\it IRAS} 12, 25, 60, and 100 $\mu$m fluxes measured three times in March 1983 and 
the differences in the individual measurements were consistent with their
uncertainties. The {\it IRAS} Point Source Catalog measurements are given in Table \ref{tab:dirbe}.

The {\it IRAS} Low Resolution Spectrometer (LRS) spectra have a resolution of $R\sim 20$ and 
cover 7.5-22.5\,$\mu$m. The $\alpha$~Ori LRS spectra were combined with the 
correction factors from \cite{cohen_etal92} and the system responses
from the \citet{iras} to derive a color-corrected 12\,$\mu$m flux of $3393\pm 136$~Jy. 
For the 25, 60 and 100\,$\mu$m fluxes we used the non-contemporaneous {\it ISO} SWS and LWS 
spectral shape to derive color-corrected fluxes which are given in Table
\ref{tab:dirbe}. 

\subsection{Kuiper Airborne Observatory}

\cite{haas_glassgold93} report the 1992 January 16 detection of
\up[\ion{Si}{2}\up] 34.81\,$\mu$m and \up[\ion{O}{1}\up] 63.18\,$\mu$m emission lines 
using the cryogenic grating spectrometer (CGS) on-board the {\it KAO} with resolutions
of $R\simeq 2900$ (44\arcsec\, aperture) and $R\simeq 3700$ (34\arcsec\, aperture), respectively. 
The measured line fluxes are: \up[\ion{Si}{2}\up] $=0.94\pm 0.37\times 10^{-18}\>{\rm W\>cm}^{-2}$ and 
\up[\ion{O}{1}\up] $=2.37\pm 0.21\times 10^{-18}\>{\rm W\>cm}^{-2}$. 
They also measured nearby continuum fluxes of $724\pm 29$\,Jy at 35\,$\mu$m, and  $96\pm 41$\,Jy at 63\,$\mu$m. 

\begin{table}[th]
\begin{center}
\caption{DIRBE and {\it IRAS} photometry: color-uncorrected and color-corrected flxues.} 

\label{tab:dirbe}
\begin{tabular}{lcccl}
\tableline\tableline
Band & Wavelength  &  Color-Uncorrected           & Color-Corrected  & $K_\lambda\>$ Spectral shape\tablenotemark{a} \\
     & ($\mu$m)        & Flux $\pm 1\sigma$ (Jy)       &  Flux $\pm 1\sigma$ (Jy) &  \\
\underline{DIRBE} & \multicolumn{3}{c}{} \\
Band 1 & 1.25     & $3.04 \pm 0.11 \times 10^{4}$ & $3.20 \pm 0.12 \times 10^{4}$ & 0.95 MARCS    \\
Band 2 & 2.2      & $3.03 \pm 0.09 \times 10^{4}$ & $3.62 \pm 0.10 \times 10^{4}$ & 0.88 MARCS    \\
Band 3 & 3.5      & $1.71 \pm 0.05 \times 10^{4}$ & $1.78 \pm 0.05 \times 10^{4}$ & 0.96 ISO-SWS  \\
Band 4 & 4.9      & $7.16 \pm 0.31 \times 10^{3}$ & $7.46 \pm 0.32 \times 10^{3}$ & 0.96 ISO-SWS  \\
Band 5 & 12       & $3.86 \pm 0.14 \times 10^{3}$ & $3.68 \pm 0.16 \times 10^{3}$ & 1.03 ISO-SWS  \\
Band 6 & 25       & $2.91 \pm 0.06 \times 10^{3}$ & $1.48 \pm 0.04 \times 10^{3}$ & 1.95 ISO-SWS  \\
Band 7 & 60       & $6.29 \pm 1.02 \times 10^{2}$ & $4.31 \pm 0.70 \times 10^{2}$ & 1.46 ISO-LWS  \\
Band 8 & 100      & $1.02 \pm 2.76 \times 10^{2}$ & $0.87 \pm 2.36 \times 10^{2}$ & 1.17 ISO-LWS  \\
       &          &                               &         &              \\
\underline{\it IRAS} & \multicolumn{3}{c}{} \\
       & 12       & $4.68 \pm 0.19 \times 10^{3}$ &  $3.39 \pm 0.14 \times 10^{3}$  & 1.38 {\it IRAS} LRS  \\
       &          &                               &  $3.87 \pm 0.16 \times 10^{3}$  & 1.21 ISO-SWS         \\
       & 25       & $1.74 \pm 0.07 \times 10^{3}$ &  $1.20 \pm 0.05 \times 10^{3}$  & 1.45 ISO-SWS         \\
       & 60       & $2.99 \pm 0.21 \times 10^{2}$ &  $2.23 \pm 0.16 \times 10^{2}$  & 1.34 ISO-LWS         \\
       & 100      & $9.59 \pm 1.92 \times 10^{1}$ &  $9.13 \pm 1.83 \times 10^{1}$  & 1.05 ISO-LWS         \\\tableline
\end{tabular}
\tablenotetext{a}{ The spectral shapes adopted to find the color-corrections, $K_\lambda$.}
\end{center}
\end{table}

\begin{figure}
\epsscale{0.6}
\plotone{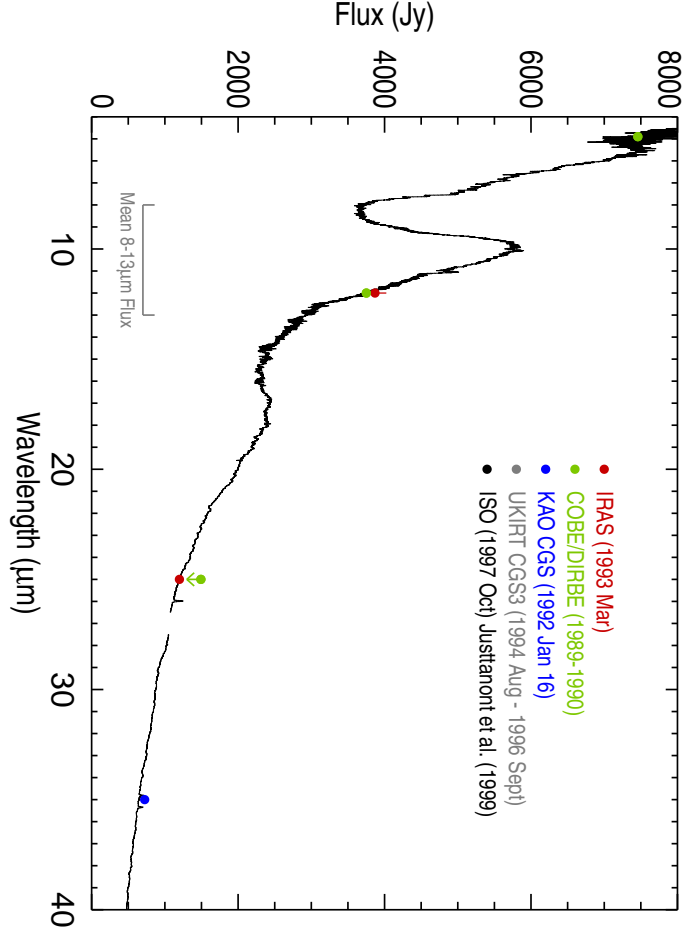}
\caption{{\it ISO} spectra and photometry of $\alpha$~Ori described in Appendix A. 
The $1\sigma$ error bars are mostly within the filled circles. The flux curve of
\cite{justtanont_etal99} is in good agreement with the {\it IRAS} fluxes and the
DIRBE 4.9 and 12\,$\mu$m fluxes. The DIRBE 25 and 60 $\mu$m fluxes are upper-limits because
of the large beam size and the presence of extended IR emission surrounding Betelgeuse. 
The {\it KAO} $3\sigma$ absolute flux uncertainty is 25\% \citep{haas_glassgold93}.
We note that the calibration of \cite{verhoelst_etal06}
appears systematically low ($> 20$\%) at these wavelengths, but these authors note that
their adopted multiplicative factors are lower than typically adopted. The mean
8-13\,$\mu$m fluxes from UKIRT CGS3 spectrophotometry between 1994 Aug and 1996 Sept
\citep{monnier_etal98} ranged from 4602-4943~Jy with 10\% uncertainty in the
absolute flux calibration. The mean 1997 8-13\,$\mu$m value derived from \cite{justtanont_etal99} is 
10\% lower at 4386~Jy.
A color version of this figure is available in the electronic edition.
\label{fig:calibrate_iso}}
\end{figure}

\subsection{Adopted Fluxes}

The IR observations discussed above were made at different epochs and
stellar variability is an important consideration.
\cite{monnier_etal98} present seven mean 8-13\,$\mu$m fluxes obtained between 
1994 August and 1996 September and the variation of these fluxes were within the 
absolute flux calibration uncertainties of 5-10\%. In the mid-1990's there is a hint that 
the 9.7\,$\mu$m silicate emission feature shifted slightly to shorter wavelengths 
than in the 1983 {\it IRAS} LRS spectra \citep{monnier_etal99}. These authors also
found that the LRS spectra are somewhat bluer than both previous and later
observations, indicating a residual miscalibration of the {\it IRAS} LRS spectra.
In light of this we redetermined the color-correction for the {\it IRAS} 12\,$\mu$m flux 
(also given in Table \ref{tab:dirbe})
using the {\it ISO}-SWS spectra and this
leads to a 15\% difference. \cite{bester_etal96} give eight 11.15\,$\mu$m fluxes obtained
between 1989 November and 1995 August and the variations of these are also consistent
with a 5\% uncertainty.  Although the observations are scarce, it appears that intrinsic star+dust 
flux variations at 12\,$\mu$m on decadal timescales are 
less than 10\%, and at 25\,$\mu$m probably less. 

The DIRBE 25\,$\mu$m and 60\,$\mu$m fluxes from 1989-1990 are significantly
larger than the 1983 {\it IRAS} fluxes, and the 25$\mu$m fluxes are
greater than the \cite{verhoelst_etal06} reduction of the
1997 {\it ISO} observations, which is expected to have a typical 
photometric uncertainty of $\sim 10$\% \citep{van_malderen_etal04}. 

The majority of the stellar wind dust emission at 25\,$\mu$m is expected to lie
within $1\arcmin$ of $\alpha$~Ori \citep{stencel_etal88}, i.e.,
well within the large DIRBE beam of $0.\degr 7\times 0.\degr 7$ and
the IRAS detectors FOV $0.75\times 4.6\arcmin$. Only a
few percent of dust emission is not collected in the {\it ISO} band 3C and 3D apertures. 
\cite{noriega_crespo97} discovered a wind-ISM bowshock located at 
5\arcmin { } from the star, and a nearby linear structure. 
The DIRBE fluxes contain contributions from this low surface intensity structured 
emission. At 60\,$\mu$m
the bowshock alone has $\sim 30$\% of the flux from the star+wind. The DIRBE 
color-corrected 60\,$\mu$m
flux $\simeq 445$\,Jy is greater than the combined star+bowshock flux
\citep{noriega_crespo97} and while the relative contribution of the
extended emission at 25\,$\mu$m is likely to be less than that at 60\,$\mu$m
it probably accounts for some of the excess DIRBE 25\,$\mu$m flux. At 12\,$\mu$m the DIRBE flux should be dominated
by the star and its dusty wind, and is consistent with the {\it IRAS} flux.
So although the DIRBE absolute photometry is very good, the large beam size makes
it less suitable to normalize the {\it ISO} spectra for $\lambda > 12\,\mu$m. 

A comparison of the 4.9 and 12\,$\mu$m DIRBE, the 12 and 25\,$\mu$m {\it IRAS},
and the 35\,$\mu$m {\it KAO} fluxes with the different {\it ISO} reductions
reveals a good overall agreement with \cite{justtanont_etal99} but not the
\cite{verhoelst_etal06} spectrum which is significantly lower.
Indeed \cite{verhoelst_etal06} noted that they scaled the sub-band fluxes
down significantly more than expected for $\lambda> 4\,\mu$m.
Our results suggest that the {\it ISO} calibration for bright IR sources 
requires improvement.

In summary, for the absolute flux scaling of the TEXES spectra we adopt 
the {\it ISO} SWS spectrum normalized to the mean of the 12 \& 25\,$\mu$m {\it IRAS} fluxes
and the 4.9 and 12\,$\mu$m DIRBE fluxes. We assign
an absolute flux uncertainty of 20\% to account for 
the combined stellar variability and the scatter in the
different mission normalizations. The {\it ISO} spectra scaled to these fluxes is shown
in Figure \ref{fig:calibrate_iso}.

The relative fluxes for the TEXES \up[\ion{Fe}{2}\up] lines ultimately relies
on the normalization of the slightly overlapping {\it ISO} spectra in band 3C 
($\lambda =16.5-19.5\>\mu$m) and band 3D ($\lambda= 19.5-27.3\>\mu$m),
e.g., between ($\lambda 19.37-19.57\>\mu$m) \citep{sloan_etal03}.
The difference in the band 3C and band 3D multiplicative factors
listed by \cite{verhoelst_etal06} is $\simeq 1.3\%$, we therefore
adopt a 2\% relative flux error in the continua near 
the [\ion{Fe}{2}] 17.98\,$\mu$m and 24.52\,$\mu$m lines. 

\subsection{Putting it all together}

The inner dust radius is measured to be $\sim 1$\arcsec\,\citep{danchi_etal94} 
and is comparable to the TEXES slit width in the dispersion direction.
The observed TEXES spectra thus suffer different slit losses for the
point source photospheric and chromospheric emission and for the more 
diffuse extended dust continuum emission. 
To place the TEXES spectra onto an absolute flux scale we must first apply 
corrections for the emission not transmitted through the 
$2\times 17\arcsec$ slit and telescope and instrument losses. 
To estimate the separate slit losses of star and dust
we use the silicate dust specific intensity model from HBL01 
and convolve the resulting sky-image with a Gaussian to represent 
a combination of seeing, diffraction and pointing jitter, whose
width is estimated by the recorded TEXES spatial profile:
FWHM$ \sim 2.6\arcsec$ at 24.5\,$\mu$m, and 
FWHM$ \sim 2.1\arcsec$ at 17.9\,$\mu$m.

The flux recorded by TEXES
\begin{equation}
F_{TEXES} = A_{loss}\left[ C_\ast F_\ast + C_{dust} F_{dust}\right]
\end{equation}
where  $A_{loss}$ is a multiplicative factor to account for combined 
telescope and instrument light losses not already corrected for
in the radiometric flat-field procedure, and $C_\ast$ and $C_{dust}$ are the fractions of the
total star and dust flux that pass through the slit, respectively. These are calculated assuming
the HBL01 sky intensity model, i.e., at 24.5\,$\mu$m: $C_\ast=0.65$ and $C_{dust}=0.31$. 

We assume that the {\it ISO} aperture records the total flux from the system, i.e.,
the star and wind emission but not the bowshock emission then
\begin{equation}
F = F_\ast + F_{dust}.
\end{equation}

At 24.5\,$\mu$m, 17.9\,$\mu$m and 11.15\,$\mu$m the ratio of dust to star
emission derived from the HBL01 model is $F_{dust}/F_\ast=1.6, 1.4, 0.7$, respectively. 
Coefficients $A_{loss}$ can then be found and the TEXES spectra corrected and scaled to the 
{\it ISO} spectrum (which is scaled to the DIRBE and {\it IRAS} fluxes) using the four TEXES orders recorded 
at each wavelength setting. The resulting TEXES spectra for $\alpha$~Ori are shown in Fig.~\ref{fig:texes_format}.

We note that the $6\arcsec$ nod of the star along the slit
followed by a subtracted image which cancels the sky noise 
clips some of the dust emission which is present at $3\arcsec$.
The spatial profile shows some emission beyond the Gaussian core 
and this is also predicted from the HBL01 dust model.
While this does not affect the present emission line analysis 
we estimate that the flux measurement procedure underestimates the 
total flux by $\sim 5-8\%$.

\section{Composite Model Atmosphere}

To determine the formation radii of Betelgeuse's mid- and far-IR emission
lines discussed in \S5 requires a comprehensive model that encompasses the chromosphere, inner wind, and CSE.
Currently no such comprehensive models exist. Models
do exist for the inner region, (HBL01: \citealt{harper_etal01}), and the CSE 
(RG91: \citealt{rodgers_glassgold91}), 
and here we describe a composite dynamic and thermodynamic 1-D model 
that utilized these models,  and interpolates between them.

The HBL01 model was based on the {\it Hipparcos} $\alpha$~Ori distance of 131~pc, but 
fortuitously the revised distance 
of $197\pm 45$~pc \citep{harper_etal08} is also that originally adopted in RG91 (200~pc). 
We therefore take Rodger \& Glassgold's stellar parameters as our nominal values:
$R_\ast =1078 \,R_\odot$, and $\phi_\ast=50$\,mas.  In \cite{harper_etal01}
the Infrared Spatial Interferometer 11.15\,$\mu$m angular diameter of 56\,mas \citep{bester_etal96} 
was adopted, but it now
appears that this may be an over-estimate of the photospheric size \citep{perrin_etal07} 
and the RG91 value is probably closer to the actual value.

\subsection{Thermal Structure}

The HBL01 model was based on angular resolved radio emission and
the old {\it Hipparcos} distance of 131\,pc. Although this model is insensitive to
the uncertain angular diameter of the photosphere, the thermodynamic 
properties must be scaled to the improved distance estimate of 197\,pc 
\citep{harper_etal08}. The radio
interferometry of \cite{lim_etal98} essentially measures the angular
distribution of the specific intensity.

If we consider the model in terms of the normalized radial distance $Z$, where $R=Z R_\ast$, 
then $T_{gas}(Z)$ remains unchanged but $R_\ast$ is now a factor of 
197/131 larger. The angular
radio brightness distribution also requires that the optical depth
through the atmosphere remains unchanged, so that 
$\tau \propto n_e n_H dR$ is constant. Therefore to satisfy this constraint the particle densities 
each scale as $1/\sqrt{197/131}$.

The single component 1-D temperature structure derived from the radio
represents a complicated averaging of the electron temperatures of the
hot chromospheric plasma and cool wind plasma. We expect that the 
filling factor of the hot plasma decreases with increasing radius, 
so the bulk of the plasma is cooler than inferred from the radio. 
For the calculation of the mid-IR emission we adopt a lower temperature 
distribution that joins the HBL01 model at $Z=7$.

When the RG91 models were constructed it was widely believed, on the basis of
theoretical grounds and semi-empirical models based on spatially unresolved 
data, that the inner wind had warm chromospheric temperatures and RG91 
adopted a nominal inner boundary condition (BC) of $T_{gas}=8000$~K at $Z=3$.
They also provided variational calculations for the temperature structures resulting from 
different mass-loss rates and where the inner BC was set to $T_{gas}=4000$~K. In contrast $T_{gas}=2764$~K for HBL01. The RG91 
models with different inner temperature BC's have similar shapes and
smoothly converge to join at 30$R_\ast$. Interior to $Z=7$ we take the temperature structure from HBL01 
which is constrained by the long-wavelength radio observations, and exterior to that the temperature structure is
obtained by extrapolating on the difference between the RG91 models with inner BC of
$T_e=4000$~K and 10,000~K. 

\begin{figure}
\epsscale{0.8}
\plotone{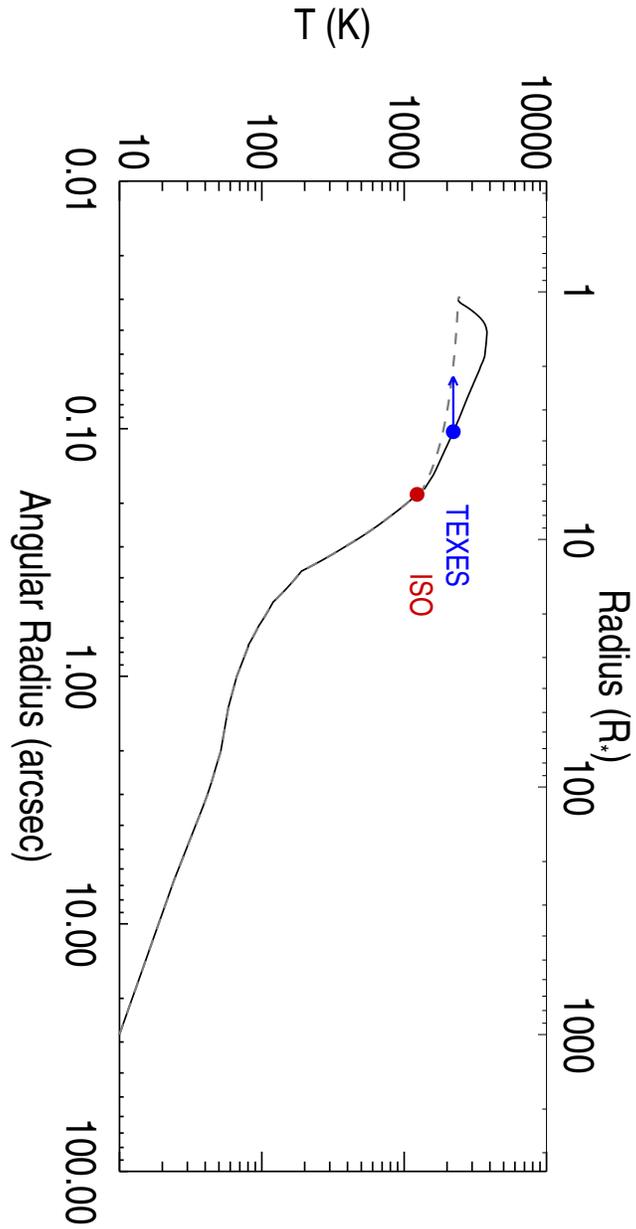}
\caption{Composite temperature structure for $\alpha$~Ori's chromosphere, inner
wind and CSE. The inner region (solid line) is based on a distance-scaled HBL01 
spatially-extended semi-empirical model, and the CSE is based on a
model of \cite{rodgers_glassgold91}. The dash line is a schematic
representation of the temperature of the dominant plasma (see text for details.)
\label{fig:te}}
\end{figure}

The CSE temperature structures are less sensitive to differences in mass-loss rates: the RG91 value of 
$3.5\times 10^{-6}\>{\rm M}_\odot{\rm yr}^{-1}$  
(assuming a mean mass per hydrogen nuclei $\Sigma=1.4$ with 
$V_{wind}=10\>{\rm km\>s}^{-1}$) is similar to  the rescaled value of HBL01 (see below). 
The adopted composite temperature structures are shown in Figure \ref{fig:te}. The 
solid line includes contribution from hot plasma while the
dashed line is a schematic representation of the temperature of the
cool wind.

\subsection{Velocity Fields}

A detailed description of the run of the mean outflow velocity, $V(R)$, and 
fluctuations about this value, $V_{turb}(R)$, are important for both 
emission line profile calculations and for determining the photon escape probability 
which enters into the flux emitted by the envelope. $V_{turb}(R)$ which enters into the line profile
calculations is assumed to be
random and isotropic (in the absence of any other knowledge) and
given by $V_{turb}(R)= \sqrt{V_{therm}^2 + V_{non-therm}^2}$, i.e., it includes
the thermal motion of the diagnostic species combined with, as yet unidentified,
non-thermal motions. Here $V_{non-therm}$ dominates.

\subsubsection{Mean Radial Outflow Velocity}

\cite{goldberg79} discussed the empirical wind velocity constraints from
atomic and molecular absorption lines. There are two clearly identified
velocity features and a dynamic region which may be physically
distinct. The S2 shell which is narrow and discrete in velocity-space [$V_{turb} \le 1\,
{\rm km\>s}^{-1}$ and $V(S2) \simeq 17\,{\rm km\>s}^{-1}$]
is observed in both absorption and 
scattered emission from atomic lines \citep{mauron90}, in 
absorption in the CO 4.6\,$\mu$m fundamental 
band \citep{bernat_etal79}, and in emission in millimeter CO rotational emission.
In K~I the S2 shell is seen out to a radius of 50\arcsec\, \citep{plez_lambert02}
with an inner edge of $\sim 7$\arcsec\, \citep{mauron90}. However, the similarity of the shape of 
CO mm-radio emission profile in single-dish observations with different beam-sizes 
suggests the outer edge of the CO is $<12$\arcsec, e.g., \cite{huggins_etal94},  
and from modeling considerations an inner edge of 2.5\arcsec\,
is plausible \citep{huggins87}. The spatial extent of different diagnostics is unlikely to be
identical because of differing ionization balances in the extended
envelope. We shall find that the atomic and singly ionized CSE emission lines are
formed within these inner radii and are likely to be characteristic of the S1 
shell [$V_{turb} \sim 4 \,{\rm km\>s}^{-1}$ and $V(S1) \simeq 10\,{\rm km\>s}^{-1}$]. 
S1 is observed in P-Cygni profiles and is blended with the photospheric absorption line, 
which may lead to a small over-estimate of the radial velocity of the shell. While S1 has not yet been 
spatially resolved in mm-radio CO observations, which show that it lies within the S2 shell \citep{harper_etal09}, 
the S1 shell has been resolved in the photospheric scattered CO 4.6\,$\mu$m fundamental band
\citep{smith_etal09}.

Dynamic flow features have been observed in ultraviolet (UV) Fe~II line profiles
\citep{boesgaard_magnan75}, and \cite{carpenter84} schematically mapped out 
the radial wind velocity using observations from the {\it International Ultraviolet
Explorer}. The wind acceleration is more seen clearly in the Goddard High Resolution
Spectrograph (GHRS) spectra studied by \cite{carpenter_robinson97} (see their Figure 6). The
line profiles map the flows close to the star and this acceleration region may be part
of S1 structure.  At 1-1.5\arcsec {} CO 4.6\,$\mu$m wind scattering observations from
Phoenix on Gemini-S (\citealt{harper_etal09, smith_etal09}) reveal that the wind
has the velocity of the S1 shell. The emission for most CSE lines (except carbon) is formed 
interior to 1\arcsec\ {} so
we will adopt a wind model that reaches a terminal speed of
$\sim 10\>{\rm km\>s}^{-1}$ at 1\arcsec {} and extended out beyond 5\arcsec. Given that it is
unlikely that there is a smooth transition in the wind properties between the
S1 and S2 material this is a reasonable procedure, until further
spatial information is obtained in this interesting region. 

We estimate the run of wind velocity close to the star using the new temperature structure to locate 
the normalized radius $(Z)$ where the absorption minima $(V_{abs})$ of UV GHRS \ion{Fe}{2} wind features have a 
radial optical depth $\tau=1$. We then select a wind velocity profile that approximates a range 
of \ion{Fe}{2} $Z-V_{abs}$ values and that is also the solution to the
constant pressure wind equation \citep{brandt70},
\begin{equation}
{V^2\over{V_{crit}^2}} - \ln{ V^2\over{V_{crit}^2} } = 4\left[ \ln{ R\over{R_{crit}}} + {R_{crit}\over{R}} \right] - 3. 
\end{equation}
The velocity profile is the solution of this transcendental equation, 
and is defined by $V_{crit}=2.5\>{\rm km\>s}^{-1}$ and $Z_{crit}=2.75$. We also
limit the wind speed not to exceed $10\>{\rm km\>s}^{-1}$ at large distance.

\subsubsection{Turbulence Velocities}

For the radial distribution of turbulence, we note that the
studies of eclipsing binaries reveal that the turbulent
velocities in the chromosphere and inner wind are 
typically $\simeq 1.5\times$ hydrogen sound speed (\citealt{eaton93, baade_etal96}), but see also
\cite{kirsch_etal01}. These small scale motions are probably related to the mass-loss mechanisms and may reflect
MHD waves, e.g., see \cite{jordan86}. 
Here we adopt $\simeq 1.5\times$ hydrogen sound speed, namely  
\begin{equation}
V_{non-therm}\left(R\right)=0.19\sqrt{T_{gas}\left(R\right)}\>\>{\rm km\>s}^{-1}.
\end{equation}
We note that this gives $V_{turb}\simeq V_{non-therm} \sim 12\>{\rm km\>s}^{-1}$ near the
base of the wind in agreement with $\alpha$~Ori's 
\up[\ion{Fe}{2}\up] line widths. 
In this model at $Z\simeq 9$ the wind and non-thermal turbulence velocities are approximately equal.

\subsection{Density Structure}

For the hydrogen densities we join the distance-scaled HBL01 model with the RG91 model.
The densities in the RG91 model are $\propto \dot{M}/V_\infty$ and they originally adopted a terminal wind speed
of $V_\infty=16\>{\rm km\>s}^{-1}$ (i.e., the S2 velocity), however, it now appears that the 
appropriate wind speed for the 
most of the CSE line formation is 
$10\>{\rm km\>s}^{-1}$ \citep{haas_glassgold93}. To maintain the same density structure with this 
lower 
velocity requires lowering the original RG91 mass-loss rate of $\dot{M}=5.6\times 10^{-6}\>{\rm M}_\odot{\rm yr}^{-1}$
to $\dot{M}=3.5\times 10^{-6}\>{\rm M}_\odot{\rm yr}^{-1}$. 
We define the hydrogen density for $Z > 7$ assuming
\begin{equation}
n_H = {\dot{M}\over{4\pi R^2 m_H \Sigma V(R)}}.
\label{eq:mcons}
\end{equation}
The densities are therefore increased over the constant wind velocity limit. At $Z=7$ the inner densities are a 
factor $\sim 2$ larger than implied by Eq \ref{eq:mcons} so the two have a simple join. 
The mass-loss rate implied by the UV Fe~II lines with this new density structure is 
$4.8 \pm 1.3 \times 10^{-6}\,{\rm M}_\odot{\rm yr}^{-1}$ which is consistent with the lower velocity RG91 value.

\subsection{Inhomogeneities}

The 1-D thermal structure derived from radio interferometry represents a mean 
value of the different structures that co-exist at a given stellar radius. Near the
temperature peak a crude estimate gives an area filling factor of hot chromospheric material
of $A_{chrom}\left(Z\right) \sim 1/3$ \citep{harper_etal01}, while by $Z\sim 3$ the filling factor is 
much smaller \citep{harper_brown06}. The filling factor of the hot plasma is expected
to continue to decrease with increasing radius. The temperature of the bulk plasma in the 
region that encompasses the hot chromosphere might then be lower than the
HB01 model, and an alternate schematic model is shown as a dash line in Fig.~\ref{fig:te},
and this is adopted in \S5 to calculate the contribution functions.


\begin{thebibliography}{}

\bibitem[Aannestad(1973)]{aannestad73} Aannestad, P. 1973, \apjs, 25, 223

\bibitem[Airapetian etal.(2000)]{airapetian_etal00} Airapetian, V. S., Ofman, L., 
         Robinson, R. D., Carpenter, K. G., \& Davila, J. 2000, \apj, 528, 965

\bibitem[Aldenius \& Johansson(2007)]{aldenius_johansson07} Aldenius, M. 
         \& Johansson, S. 2007, \aap, 467, 753

\bibitem[Altenhoff et al.(1994)]{altenhoff_etal94} Altenhoff, W. J., Thum, C., \&
         Wendker, H. J. 1994, \aap, 281, 161

\bibitem[Aoki et al.(1998)]{aoki_etal98} Aoki, W., Tsuji, T., \& Ohnaka, K. 
         1998, \aap, 333, L19

\bibitem[Ayres(2002)]{ayres02} Ayres, T. R. 2002, \apj, 460, 1042

\bibitem[Baade et al.(1996)]{baade_etal96} Baade, R., Kirsch, T., Reimers, D.,
         Toussaint, F., Bennett, P. D., Brown, A., \& Harper, G. M. 1996, 
         \apj, 466, 979

\bibitem[Bahcall \& Wolf(1968)]{bahcall_wolf68} Bahcall, J. N. \& Wolf, R. A. 
         1968, \apj, 152, 701 

\bibitem[Barbier-Brossat \& Fignon(2000)]{barbier-brossat_fignon00} Barbier-Brossat, M. \&
         Fignon, P. 2000, \aaps, 142, 217

\bibitem[Barlow(1999)]{barlow99} Barlow, M. J. 1999, IAU Symp. 191, 
         Eds. T. Le Bertre, A. L\`ebre, \& C. Waelkens, p. 353

\bibitem[Bernat(1981)]{bernat81} Bernat, A. P. 1981, \apj, 246, 184

\bibitem[Bernat et al.(1979)]{bernat_etal79} Bernat, A. P., Hall, D. N. B.,
         Hinkle, K. H., \& Ridgway, S. T. 1979, \apj, 233, L135

\bibitem[Bester et al.(1996)]{bester_etal96} Bester, M., Danchi, W. C.,
         Hale, D., Townes, C. H., Degiacomi, C. G., M\'ekarnia, \& 
         Geballe, T. R. 1996, \apj, 463, 336

\bibitem[Boesgaard \& Magnan(1975)]{boesgaard_magnan75} Boesgaard, A. \&
         Magnan, P. 1975, \apj, 198, 369

\bibitem[Boggess et al.(1992)]{boggess_etal92} Boggess, N. W., et al. 
         1992, \apj, 397, 420

\bibitem[Brandt(1970)]{brandt70} Brandt, J. C., 1970, Introduction to the Solar Wind,
Freeman San Franscico (p. Eq. 3.13)

\bibitem[Brown \& Evenson(1995)]{brown_evenson95} Brown, J. M. \&
         Evenson, K. M. 1995, \apj, 441, L97

\bibitem[Carpenter(1984)]{carpenter84} Carpenter, K. G. 1984, \apj, 285, 181

\bibitem[Carpenter \& Robinson(1997)]{carpenter_robinson97} Carpenter, K. G.
         \& Robinson, R. D. 1997, \apj, 479, 970

\bibitem[Carpenter et al.(1995)]{carpenter_etal95} Carpenter, K. G.,
         Robinson, R. D., \& Judge, P. G. 1995, \apj, 444, 424

\bibitem[Carpenter et al.(1991)]{carpenter_etal91} Carpenter, K. G.,
         Robinson, R. D., Wahlgren, G. M., Ake, T. B., Ebbets, D. C.,
         Linsky, J. L., Brown, A., \& Walter, F. M. 1991, \apj, 377, L45 

\bibitem[Carr et al.(2000)]{carr_etal00} Carr. J. S., Sellgren, K.,
         \& Balachandran, S. C. 2000, \apj, 530, 307

\bibitem[Castor(1970)]{castor70} Castor, J. I. 1970, \mnras, 149, 111

\bibitem[Castro-Carrizo et al.(2001)]{castro_carrizo_etal01} Castro-Carrizo, A., 
         Bujarrabal, V., Fong, D., Meixner, M.,
         Tielens, A. G. G. M., Latter, W. B., \& Barlow, M. J. 2001,
         \aap, 367, 674

\bibitem[Cohen et al.(1992)]{cohen_etal92} Cohen, M., Walker, R. G., \&
         Witteborn, F. C. 1992, \aj, 104, 2030

\bibitem[Crowley et al.(2008)]{crowley_etal08} Crowley, C., Espey, B. R.,
         McCandliss, S. R. 2008, \apj, 675, 711

\bibitem[Danchi et al.(1994)]{danchi_etal94} Danchi, W. C., Bester, M.,
        Degiacomi, C. G., Greenhill, , L. J., \& Townes, C. H. 1994, 
        \aj, 107, 1469

\bibitem[David \& Papoular(1990)]{david_papoular90} David, P. \& Papoular, R.
         1990, \aap, 237, 425

\bibitem[Dyck et al.(1996)]{dyck_etal96} Dyck, H. M., Benson, J. A., Van Belle, G. T.,
         \& Ridgeway, S. T. 1996, \aj, 111, 1705

\bibitem[Eaton(1993)]{eaton93} Eaton, J. A. 1993, \apj, 404, 305

\bibitem[ESA(1997)]{esa97} The {\it Hipparcos and Tycho} Catalogues,
         ESA SP-1200

\bibitem[Fuhr \& Wiese(2006)]{fuhr_wiese06} Fuhr, J. R. \& Wiese, W. L.
         2006, J. Phys. Chem. Ref. Data, 35, No. 4

\bibitem[Garstang(1962)]{garstang62} Garstang, R. H. 1962, \mnras, 124, 321

\bibitem[Goldberg(1979)]{goldberg79} Goldberg, L. 1979, QJRAS, 20, 361

\bibitem[Gray(2000)]{gray00} Gray, D. F. 2000, \apj, 532, 487

\bibitem[Gray(2001)]{gray01} Gray, D. F. 2001, \pasp, 113, 1378

\bibitem[Gray(2008)]{gray08} Gray, D. F. 2008, \aj, 135, 1450

\bibitem[Haas \& Glassgold(1993)]{haas_glassgold93} Haas, M. R. \& 
          Glassgold, A. E. 1993, \apj, 410, L111

\bibitem[Haas et al.(1995)]{haas_etal95} Haas, M. R., Glassgold, A. E.,
         \& Tielens, A. G. G. M. 1995, in Airborne Astronomy Symposium on
          Galactic Ecosystem, Eds. M. R. Haas, J. A. Davidson, \&
          E. F. Erikson, ASP Conf. Ser. Vol. 73, p. 397

\bibitem[Harper(1988)]{harper88} Harper, G. M. 1988, D. Phil Thesis, University of Oxford, England

\bibitem[Harper(2001)]{harper01} Harper, G.M. 2001, in Cool Stars, Stellar Systems, and the Sun, 11th Cambridge Workshop, 
          eds. R. J. Garcia Lopez, R. Rebolo \& M. R. Zapatero Osorio, ASP Conf Ser. 223, p. 368

\bibitem[Harper et al.(2005)]{harper_etal05} Harper, G. M., Brown, A., Bennett, P. D.,
         Baade, R., Wlader, R., Hummel, C. A. 2005, \aj, 129, 1018

\bibitem[Harper \& Brown(2006)]{harper_brown06} Harper, G. M. \& 
         Brown, A. 2006, \apj, 646, 1179 

\bibitem[Harper et al.(2008)]{harper_etal08} Harper, G. M., 
         Brown, A., \& Guinan, E. F., 2008, \aj, 135, 1430

\bibitem[Harper et al.(2001)]{harper_etal01} Harper, G. M., Brown, A., \& 
         Lim, J. 2001, \apj, 551, 1073 [HBL01]

\bibitem[Harper et al.(1995)]{harper_etal95} Harper, G. M., Wood, B. E.,
         Linsky, J. L., Bennett, P. D., Ayres, T. R., \& Brown, A. 1995, 
         \apj, 452, 407  

\bibitem[Harper(2009)]{harper09} Harper, G. M. 2009, in Proceedinds of ``Hot And Cool:
         Bridging Gaps in Massive Star Evolution'', ASP COnf. Ser., In Prep

\bibitem[Harper et al.(2009)]{harper_etal09} Harper, G. M., Carpenter, K. G., 
         Ryde, N., Smith, N., Brown, J., Brown, A., \& Hinkle, K. H. 2009, 
         in Cool Stars, Stellar Systems, and the Sun, 15th Cambridge Workshop, 
         AIP Conference Proceedings, ed. E. Stempels, In Press

\bibitem[Hartmann \& Avrett(1984)]{hartmann_avrett84} Hartmann, L. \&
         Avrett, E. H. 1984 \apj, 284, 238

\bibitem[Hartmann et al.(1981)]{hartmann_etal81} Hartmann, L., Dupree, A. K., \&
         Raymond, J. C. 1981 \apj, 246, 193

\bibitem[Hartmann \& MacGregor(1980)]{hartmann_macgregor80} Hartmann, L. \&
         MacGregor, K. B. 1980 \apj, 242, 260

\bibitem[Hauser et al.(1998)]{hauser_etal98} Hauser, M. G., Kelsall, T., 
         Leisawitz, D., \& Weiland, J. 1998, {\it COBE} Diffuse Infrared 
         Background Experiment (DIRBE) Explanatory Supplement, Vers. 2.3
         (Greenbelt: NASA)

\bibitem[Hebden et al.(1987)]{hebden_etal87} Hebden, J. C., Eckart, A.,
         \& Hege, E. K. 1987, \apj, 314, 690


\bibitem[Hjellming \& Newell(1983)]{hjellming_newell83}  Hjellming, R. M. \&
         Newell, R. T. 1983, \apj, 275, 704


\bibitem[Hollenbach \& McKee(1989)]{hollenbach_mckee89} Hollenbach, D. \&
         McKee, C. F. 1989, \apj, 342, 306

\bibitem[Holzer et al.(1983)]{holzer_etal83} Holzer, T. E., Fl\aa, T., \&
         Leer, E. 1983, \apj, 275, 808

\bibitem[Holzer \& MacGregor(1985)]{holzer_macgregor85} Holzer, T. E. \&
         MacGregor, K. B. 1985, in {\it  Mass Loss from Red Giants}, eds.
         M. Morris \& B. Zuckerman, Reidel Publishing Company, p. 229

\bibitem[Huggins(1987)]{huggins87} Huggins, P. J. 1987, \apj, 313, 400

\bibitem[Huggins et al.(1994)]{huggins_etal94} Huggins, P. J., Bachiller, R., 
         Cox, P., \& Forveille, T. 1994, \apj, 424, L127

\bibitem[Hummer(1981)]{hummer81} Hummer, D. G. 1981, JQSRT, 26, 187

\bibitem[Hummer \& Storey(1992)]{hummer_storey92} Hummer, D. G. \&
         Storey, P. J. 1992, \mnras, 254, 277

\bibitem[{\it IRAS} Explanatory Supplement(1988)]{iras} 
  IRAS Catalogs and Atlases, Volume 1, ed. C. Beichman et al., NASA RP-1190, 
  (Washington, DC: US Government Printing Office.)

\bibitem[Jones(1928)]{jones28} Jones, H. S. 1928, \mnras, 88, 660

\bibitem[Jordan(1986)]{jordan86} Jordan, C. 1986, Irish Astron. J., 17, 227

\bibitem[Justtanont et al.(1999)]{justtanont_etal99} Justtanont, K., 
         Tielens, A. G. G. M., de Jong, T., Cami, J., Waters, L. B. F. M.,
         \& Yamamura, I. 1999, \aap, 345, 605

\bibitem[Kelly \& Lacy(1995)]{kelly_lacy95} Kelly, D. M. \& Lacy, J. H.
         1995, \apj, 454, L161

\bibitem[Kirsch et al.(2001)]{kirsch_etal01} Kirsch, T., Baade, D., \& 
         Reimers, D. 2001, \aap, 379, 925

\bibitem[Kunasz \& Hummer(1974)]{kunasz_hummer74} Kunasz, P. B. \& Hummer, D. G.
         1974, \mnras, 166, 19

\bibitem[Lacy et al.(1989)]{lacy_etal89} Lacy, J. H., Achtermann, J. M., Bruce, D. E.,
         Lester, D. F., Arens, J. F., Peck, M. C., \& Gaalema, S. D. 1989, \pasp, 101, 1166

\bibitem[Lacy et al.(2002)]{lacy_etal02} Lacy, J. H., Richter, M. J., 
         Greathouse, T. K., Jaffe, D. T., \& Zhu, Q. 2002, \pasp,
         114, 153

\bibitem[Lambert et al.(1984)]{lambert_etal84} Lambert, D. L., Brown, J. A.,
         Hinkle, K. H., Johnson, H. R. 1984, \apj, 284, 223 

\bibitem[Lim et al.(1998)]{lim_etal98} Lim, J., Carilli, C. J., 
         White, S. M., Beasley, A. J., \& Marson, R. G. 
         1998, \nat, 392, 575

\bibitem[Lobel \& Dupree(2000)]{lobel_dupree00} Lobel, A. \& Dupree, A. K.
         2000, \apj, 545, 454

\bibitem[Lobel \& Dupree(2001)]{lobel_dupree01} Lobel, A. \& Dupree, A. K.
         2001, \apj, 558, 815

\bibitem[Marsh et al.(2001)]{marsh_etal01} Marsh, K. A., Bloemhof, E. E., 
         Koerner, D. W., \& Ressler, M. E. 2001, \apj, 548, 861

\bibitem[Mamon et al.(1988)]{mamon_etal88} Mamon, G. A., Glassgold, A. E., \& Huggins, P. J.
         1988, \apj, 328, 797

\bibitem[Mauron(1990)]{mauron90} Mauron, N. 1990, \aap, 227, 141

\bibitem[Mermilliod et al.(2008)]{mermilliod_etal08} Mermilliod, J. C., Mayor, M., \&
         Udry, S. 2008, \aap, 485, 303

\bibitem[Mihalas(1978)]{mihalas78} Mihalas, D. 1978 Stellar Atmopsheres, W. H. Freeman \& Company,
 2nd Edition, p. 376.

\bibitem[Monnier et al.(1998)]{monnier_etal98} Monnier, J. D., Geballe, T. R.,
         \& Danchi, W. C. 1998, \apj, 502, 833

\bibitem[Monnier et al.(1999)]{monnier_etal99} Monnier, J. D., Geballe, T. R.,
         \& Danchi, W. C. 1999, \apj, 521, 261

\bibitem[Newell \& Hjellming(1982)]{newell_hjellming82} Newell, R. T. \&
         Hjellming, R. M. 1982, \apj, 263, L85

\bibitem[Noriega-Crespo et al.(1997)]{noriega_crespo97} Noriega-Crespo, A.,
         van Buren, D., Cao, Y., \& Dgani, R. 1997, \aj, 114, 837

\bibitem[Nussbaumer \& Storey(1988)]{nussbaumer_storey88} Nussbaumer, H. \&
         Storey, P. J. 1988, \aap, 193, 327

\bibitem[Perrin et al.(2007)]{perrin_etal07} Perrin, G., et al.
         2007, \aap, 474, 599

\bibitem[Plez \& Lambert(2002)]{plez_lambert02} Plez, B. \& Lambert, D. L.
         2002, \aap, 386, 1009

\bibitem[Pradhan \& Zhang(1993)]{pradhan_zhang93} Pradhan, A. K. \&
         Zhang, H. L. 1993, \apj, 409, L77

\bibitem[Quinet, Le Dourneuf \& Zeippen(1996)]{quinet_etal96} Quinet, P.,
         Le Dourneuf, M., \& Zeippen, C. J. 1996, \aaps, 120, 361

\bibitem[Ramsbottom et al.(2007)]{ramsbottom_etal07} Ramsbottom, C. A.,
         Hudson, C. E., Norrington, P. H., \& Scott, M. P. 2007,
         \aap, 475, 765 

\bibitem[Reimers et al.(2008)]{reimers_etal08} Reimers, D., Hagen, H.-J., Baade, R., \&
         Braun, K. 2008, \aap, submitted

\bibitem[Richter et al.(2006)]{richter_etal06} Richter, M. J., Lacy, J. H., Jaffe, D. T.,
         Mar, D. J., Goertz, J., Moller, M., Strong, S., Gretahouse, T. K. 2006, 
         SPIE, 6269, 49

\bibitem[Rodgers(1990)]{rodgers90} Rodgers, B. 1990, M,S Thesis, New York University

\bibitem[Rodgers \& Glassgold(1991)]{rodgers_glassgold91} Rodgers, B. \&
         Glassgold, A. E. 1991, \apj, 382, 606 [RG91]

\bibitem[Ryde et al.(2002)]{ryde_etal02} Ryde, Lambert, D. L., Richter, M. J., 
         \& Lacy, J. H. 2002, \apj, 580, 447

\bibitem[Ryde et al.(2006a)]{ryde_etal06a} Ryde, N., Harper, G. M.,   
         Richter, M. J., Greathouse, T. K., \& Lacy, J. H. 
         2006, \apj, 637, 1040

\bibitem[Ryde et al.(2006b)]{ryde_etal06b} Ryde, N., Richter, M. J., Harper, G. M., 
         Eriksson, K., \& Lambert, D. L. 2006, \apj, 645, 652

\bibitem[Sanford(1933)]{sanford33} Sanford, R. F. 1933, \apj, 77, 110

\bibitem[Sault, Teuben, \& Wright(1995)]{sault95}
         Sault, R. J., Teuben, P. J., \& Wright, M. C. H. 1995, in Astronomical 
         Data Analysis Software and Systems IV, eds. R. A. Shaw,
         H. E. Payne, \& J. J. E. Hayes (san Francisco: ASP), 
         ASP Conf. Ser., 77, 433

\bibitem[Scoville et al.(1993)]{Scoville93}
         Scoville, N. Z., et al. 1993, PASP, 105, 1482

\bibitem[Sigut \& Pradhan(1998)]{sigut_pradhan98} Sigut, T. A. A. \& Pradhan, A. K. 
         1998, \apj, 499, L139

\bibitem[Skinner et al.(1997)]{skinner_etal97} Skinner, C. J., Dougherty, S. M.,
         Meixner, M., Bode, M. F., Davis, R. J., Drake, S. A., Arens, J. F., 
         Jernigan, J. G. 1997, \mnras, 288, 295         

\bibitem[Skinner et al.(1990)]{skinner_etal90} Skinner, C. J., Griffin, I., \&
         Whitmore, B. 1990, \mnras, 243, 78      

\bibitem[Skinner \& Whitmore(1988)]{skinner_whitmore88} Skinner, C. J. \&
         Whitmore, B. 1988, \mnras, 235, 603

\bibitem[Sloan \& Price(1998)]{sloan_price98} Sloan, G. C. \& Price, S. D.
         1998, \apjs, 119, 141

\bibitem[Sloan et al.(2003)]{sloan_etal03} Sloan, G. C., Kraemer, K. E.,
         Price, S. D., \& Shipman, R. F. 2003, \apjs, 147, 379

\bibitem[Smith(2003)]{smith03} Smith, B. J. 2003, \aj, 126, 935    

\bibitem[Smith et al.(2004)]{smith_etal04} Smith, B. J., Price, S. D., \&
         Baker, R. I. 2004, \apjs, 154, 673

\bibitem[Smith et al.(1989)]{smith_etal89} Smith, M. A., Patten, B. M., 
         \& Goldberg, L. 1989, \aj, 98, 2233

\bibitem[Smith et al.(2009)]{smith_etal09} Smith, N., Hinkle, K. H., \& Ryde, N.
         2009, \apj, accepted.

\bibitem[Stencel, Pesce \& Hagen Bauer(1988)]{stencel_etal88}
         Stencel, R. E., Pesce, J. E., \& Hagen Bauer, W. 1988, 
         \aj, 95, 141

\bibitem[Swings \& Preston(1978)]{swings_preston78} Swings, J. P. \& 
         Preston, G. W. 1978, \apj, 220, 883

\bibitem[Van Malderen et al.(2004)]{van_malderen_etal04} Van Malderen, R.,
         Decin, L., Kester, D., Vandenbussche, B., Waelkens, C.,
         Cami, J., \& Shipman, R. F., 2004, \aap, 414, 677

\bibitem[Verhoelst et al.(2006)]{verhoelst_etal06} 
          Verhoelst, T. et al., 2006, \aap, 447, 311

\bibitem[Wischnewski \& Wendker(1981)]{wischnewski_wendker81} Wischnewski, E. \&
         Wendker, H. J. 1981, \aap, 96, 102

\end{thebibliography}
\end{document}